\documentclass{jpp}
\usepackage{graphicx}

\usepackage[utf8]{inputenc}
\usepackage[T1]{fontenc}
\usepackage{amsmath}

\shorttitle{Modulated deuteron spectra observed with the MRS at the NIF}
\title{Modulated deuteron spectra observed with the Magnetic Recoil neutron Spectrometer at the National Ignition Facility}
\shortauthor{B. Nguyen and others}

\author{
B. Nguyen\aff{1,2}
  \corresp{\email{wgn21@ic.ac.uk}},
Y. Lawrence\aff{2},
C. W. Wink\aff{2},
T. M. Johnson\aff{3},
N. Vanderloo\aff{2},
B. L. Reichelt\aff{2},
A. Hennessy\aff{4},
D. T. Casey\aff{3},
D. Schlossberg\aff{3},
N. Masters\aff{3},
J. L. Milovich\aff{3},
A. Le\aff{5},
R. S. Craxton\aff{6},
M. Gatu Johnson\aff{2}
\and
J. A. Frenje\aff{2}
}

\affiliation{
\aff{1}Imperial College London, London SW7 2AZ, United Kingdom\\
\aff{2}Massachusetts Institute of Technology, Cambridge, Massachusetts 02139, USA\\
\aff{3}Lawrence Livermore National Laboratory, Livermore, California 94551, USA\\
\aff{4}General Atomics, San Diego, California 92186, USA\\
\aff{5}Los Alamos National Laboratory, Los Alamos, New Mexico 87544, USA\\
\aff{6}Laboratory for Laser Energetics, Rochester, New York 14623, USA
}

\begin{document}

\maketitle

\begin{abstract}
The Magnetic Recoil Spectrometer (MRS) on the National Ignition Facility is used to measure the neutron spectrum from deuterium-tritium fueled inertial confinement fusion implosions via n-d elastic scattering and magnetic dispersion of recoil deuterons. From the  MRS-determined neutron spectrum, the yield ($Y_n$), apparent ion temperature ($T_i$) and areal density ($\rho R$) are determined. However, anomalous energy modulations in recoil deuterons have been observed in several high-yield indirect drive experiments ($Y_n\sim10^{16}-10^{18}$). These observations raise concerns about their potential impact on the MRS-inferred performance metrics. Analytic calculations and particle-in-cell simulations are used to examine the possible beam-plasma instabilities, which indicate the two-stream instability as the driving mechanism behind energy modulations. Based on a statistical analysis of synthetic deuteron spectra, the modulations-induced errors are found to be within the errors of the determined $Y_n$, $T_i$ and $\rho R$ values and thus do not have a significant impact on the MRS measurement. 
\end{abstract}

\section{Introduction}
Since its commissioning at the National Ignition Facility (NIF), the Magnetic Recoil Spectrometer (MRS) has been routinely used to measure neutron spectra from deuterium-tritium (DT) implosions \citep{frenje2010probing, frenje2020nuclear}. These measurements have supported a wide range of inertial confinement fusion (ICF) experiments \citep{smalyuk2014measurements, lindl2014review, zylstra2021record}, including recent campaigns leading to ignition \citep{abu2022lawson}.

\begin{figure}
    \centering
    \includegraphics[scale=0.38]{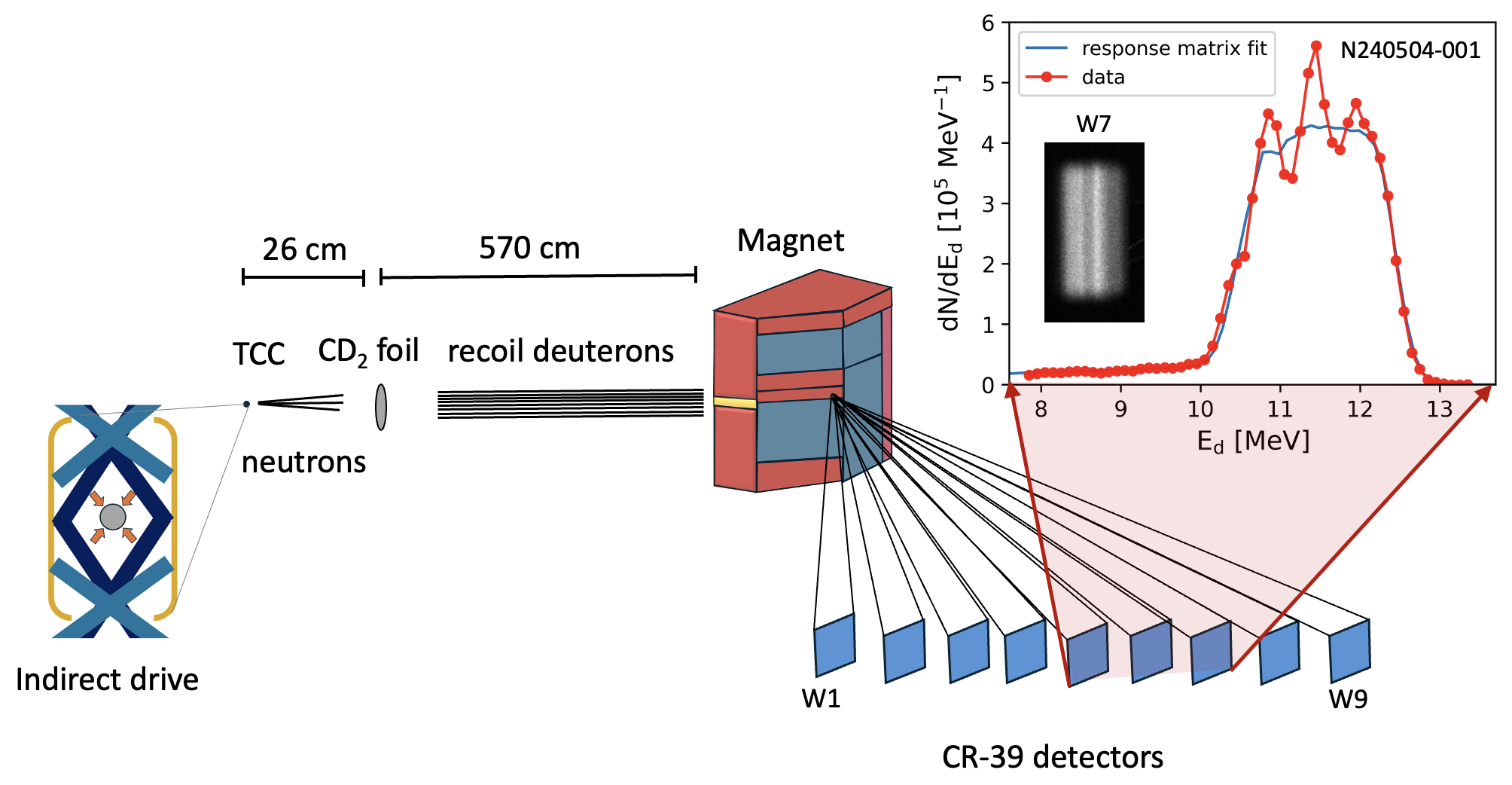}
    \caption[]{Schematic of the MRS on the NIF. The spectrum illustrated in the top-right panel shows the observed modulated deuteron spectrum (red) relative to the response matrix fit (blue) for shot  N240504-001. In the CR-39 scan image, the vertical axis corresponds to the non-dispersion direction and the horizontal axis represents the dispersion (energy) direction. The grayscale represents the number of deuteron tracks in each bin, with more white corresponding to more tracks.}
    \label{fig:MRS_schematics}
\end{figure}

As seen in figure~\ref{fig:MRS_schematics}, the MRS consists of a deuterated polyethylene ($\mathrm{CD_2}$) foil, a focusing magnet, and an array of CR-39 detectors~\citep{frenje2010probing, casey2013magnetic}. The MRS operates as follows: neutrons are converted to deuterons via neutron-deuteron (n-d) elastic scattering in the $\mathrm{CD_2}$ foil. Forward-scattered deuterons are selected by the magnet aperture, focused and dispersed onto CR-39 detectors according to their energy. Using the mapping between position and energy, the deuteron spectrum is measured. The spectrum is analyzed using a forward-fit technique, where a model neutron spectrum is convolved with the MRS response function and iteratively adjusted until a match to the measured deuteron spectrum is found. From the best-fit model, the neutron yield ($Y_n$), apparent ion temperature ($T_i$), and areal density ($\rho R$) are extracted.

Unexpected energy modulations in the MRS deuteron spectra have been observed in several high-yield NIF experiments $(Y_n \sim 10^{16}-10^{18})$. The inset of figure~\ref{fig:MRS_schematics} shows the deuteron spectrum from shot N240504-001, where the effect of modulation is the most visible. These modulations may introduce additional uncertainties to the inferred performance metrics $(Y_n,\ T_i,\ \rho R)$, and the significance of these is investigated in this paper. As the MRS is upgraded to support higher yields $Y_n \gtrsim 10^{18}$ \citep{gatu2022high}, understanding the causes of energy modulations and mitigating their impact becomes increasingly important.

Proton energy modulations have been observed in intense laser-matter interaction experiments \citep{hicks2001observations, allen2003proton, busch2004shape}. The first reported observation likely comes from the charged particle spectrometer (CPS) at the OMEGA facility, where oscillations were seen \citep{hicks2001observations}. Based on the modulation period, Hicks \emph{et al}. speculated that these structures result from ion acoustic perturbations of an expanding plasma. Similar modulations have also been observed in target normal sheath acceleration (TNSA) experiments \citep{busch2004shape,allen2003proton}. Using particle-in-cell (PIC) simulations, \cite{busch2004shape} attribute modulations to the presence of two-temperature electron distribution, whereas \cite{allen2003proton} suggest they arise from the presence of multiple heavy-ion species in an expanding plasma. 

Ion-energy modulations have also been studied through the lens of kinetic instabilities under different plasma conditions. \cite{tikhonchuk2005ion} proposed that a two-species plasma expansion can drive a two-stream instability (TSI), leading to energy and density modulations. Using PIC simulations under simplified ICF conditions ($n \sim 10^{22} \ \mathrm{cm^{-3}}$, $T_e \sim 3$~keV), \cite{lv2023ion} reproduced ion energy oscillations reminiscent of past observations \citep{hicks2001observations}. Fourier analysis of the electric field confirms that these oscillations originate from the ion-acoustic instability induced by the relative motion between counter-streaming ion species. Similar modulations have also been reported under non-ICF plasma conditions ($n \sim 10^{11} \ \mathrm{cm^{-3}}$, $T_e \sim 10$ eV). For instance, \cite{hui2019modulation} performed 3D PIC simulations of a mono-energetic proton beam propagating through a uniform plasma. They found that the initial energy peak evolves into two and later three peaks due to the TSI.

In this work, a combination of experimental data and simulations is used to constrain the foil conditions. Based on the estimated conditions, energy modulations are likely caused by the TSI driven by fast deuterons streaming through background electrons. PIC simulations reasonably reproduce the observed energy modulations and show a linear scaling with the deuteron-to-electron density ratio ($n_d/n_e$), which is stronger at higher yields. Synthetic spectra show negligible impact on MRS-inferred quantities at the current yields $Y_n\sim10^{17}-10^{18}$. However, the effect of modulations is expected to be significant at higher yields, which can be mitigated by placing the foil directly at the MRS.  

This paper is organized as follows: \S\ref{sec:exp_conditions} outlines the experimental conditions related to neutron-foil interactions and electron sources. In \S\ref{sec:instability}, analytic theory is used to identify the TSI as the mechanism behind the observed energy modulations. \S\ref{sec:pre-neutron} and~\S\ref{sec:post-neutron} present PIC simulations of shot N240504-001, where the pronounced modulations provide a clear basis for comparison. This is followed by a density scan to generalize the analysis beyond the simulated case. \S\ref{sec:sensitivity_study} concludes with a synthetic-data study and briefly discusses potential mitigation strategies. 

\section{Experimental conditions}\label{sec:exp_conditions}

\subsection{Neutron-foil interactions}

Near peak convergence of the fuel, DT fusion in the hotspot produces energetic neutrons (14.1 MeV). Their energy spectrum is Doppler-broadened by the ion temperature $T_i$, resulting in primary neutrons ($13-15$ MeV). Depending on $\rho R$, a fraction of these neutrons scatter off fuel ions and lose energy, producing down-scattered neutrons ($10-12$ MeV) \citep{frenje2013diagnosing}.
 
\begin{figure}
    \centering
    \includegraphics[scale=0.9]{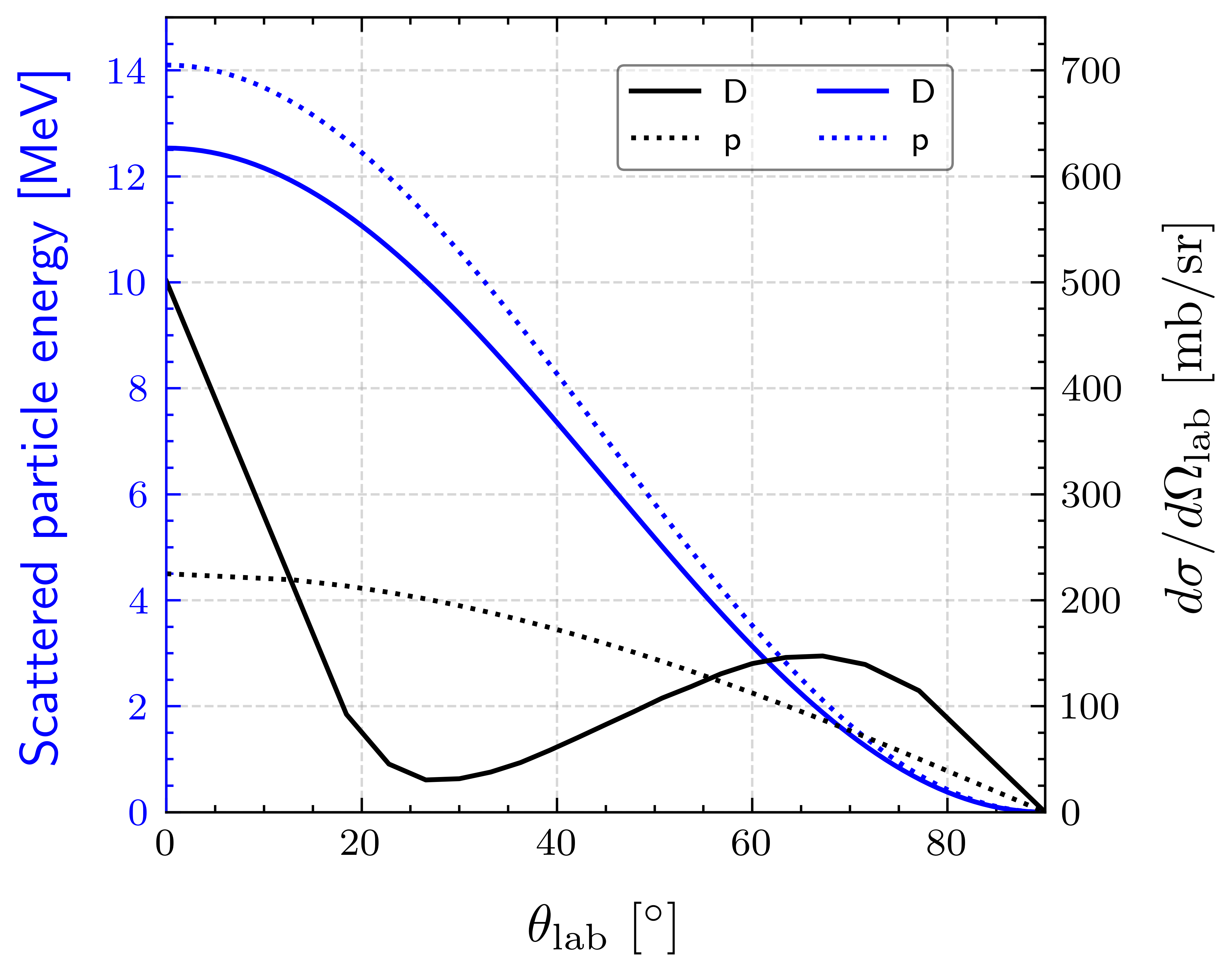}
   \caption[]{Scattered particle energy (blue) and angular distribution (black) as a function of $\theta_\text{lab}$ for deuterons (solid) and protons (dotted). The MRS aperture selects particles with $\theta_\text{lab} \sim 0$.}
  \label{fig:nd_scattering}
\end{figure}

Recoil deuterons are generated via n-d elastic scattering in a $\mathrm{CD_2}$ foil located 26 cm from the target chamber center (TCC). By conservation of momentum and energy, the deuteron energy $E_d$ is given by

\begin{equation} 
E_d = \frac{8}{9} E_n \cos^2{\theta_\text{lab}}, 
\label{eq:recoil_energy} 
\end{equation}
where $E_n$ is the incident neutron energy and $\theta_\text{lab}$ is the laboratory scattering angle \citep{frenje2013diagnosing}. The differential distribution of $\theta_\text{lab}$ is extracted from the nuclear data repository ENDF \citep{brown2018endf}. Figure~\ref{fig:nd_scattering} displays the deuteron energy and angular distribution as a function of $\theta_\text{lab}$ for $E_n=14.1$ MeV. Due to the location and geometry of the magnet aperture relative to the foil, only forward scattered deuterons ($\theta_\text{lab}\sim0$) are selected by the MRS. The spectral shape of these aperture-selected deuterons depends on the MRS response function, which was previously simulated by \cite{casey2013magnetic} using the multi-physics code Geant4 \citep{agostinelli2003geant4, allison2006geant4, allison2016recent}. 

Neutron-foil interactions can produce a range of other charged particles in addition to deuterons. To characterize the particle output, Geant4 simulations are performed, where an isotropic neutron source is placed 26 cm from a $\mathrm{CD_2}$ foil. The resulting charged particles are listed in table~\ref{tab:geant_particle_output}, where the dominant species consist of deuterons and protons (relative ratio $\sim 3:1$). The number of ejected electrons is comparatively low and insufficient to neutralize the outgoing ion population. Charge separation effects and the resulting self-consistent electric fields are not captured in Geant4 and are instead modeled using PIC simulations in \S\ref{sec:post-neutron}.

\begin{table}
    \centering
    \def~{\hphantom{0}}
    \begin{tabular}{l@{\hskip 1.5em}c@{\hskip 1.5em}c}
    \hline \hline
        Species & Normalized particle count & Mean kinetic energy [MeV] \\ [3pt]
    \hline
        Be-10   & $1.0 \times 10^{-11}$ & 2.89 \\ [2pt]
        Be-9    & $7.8 \times 10^{-10}$ & 0.71 \\ [2pt]
        C-12    & $7.1 \times 10^{-9}$  & 0.97 \\ [2pt]
        C-13    & $9.8 \times 10^{-11}$ & 1.27 \\ [2pt]
        He-3    & $5.5 \times 10^{-13}$ & 7.98 \\ [2pt]
        Li-6    & $5.5 \times 10^{-13}$ & 1.74 \\ [2pt]
        He-4    & $1.7 \times 10^{-9}$  & 1.31 \\ [2pt]
        e$^+$   & $3.9 \times 10^{-11}$ & 1.52 \\ [2pt]
        e$^-$   & $6.0 \times 10^{-10}$ & 1.66 \\ [2pt]
        p       & $1.4 \times 10^{-7}$  & 4.42 \\ [2pt]
        D       & $4.1 \times 10^{-7}$  & 6.01 \\ [2pt]
        T       & $5.5 \times 10^{-13}$ & 9.78 \\ [2pt]
    \hline \hline
    \end{tabular}
    \caption{Summary of charged particles emerging from the foil in Geant4. The particle count is normalized per source neutron per $\mu$m$\cdot$cm$^2$ (thickness-area) foil. The statistical uncertainty for deuteron and proton is $\lesssim 0.1\%$.}
    \label{tab:geant_particle_output}
\end{table}

\subsection{Sources of electrons}
 
\begin{figure}
    \centering
    \includegraphics[scale=0.4]{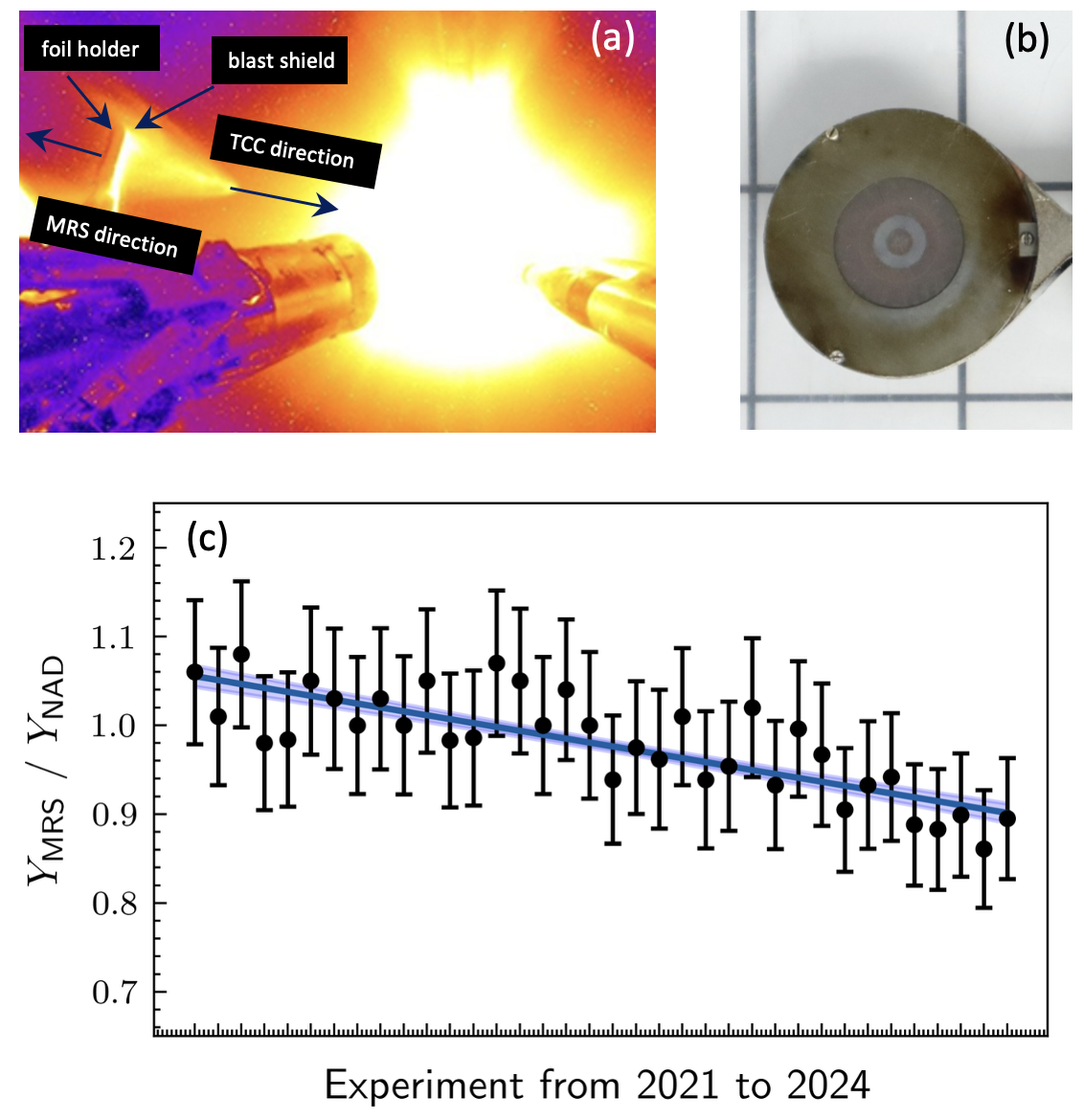}
   \caption[]{(a) Optical image showing bright conical structures around the MRS foil and blast shield. (b) Image of a partially burnt foil. (c) Ratio of the inferred yield from the MRS to the neutron activation detectors (NAD) for 36 shots fielded with the foil shown in (b).} 
  \label{fig:foil_ablation}
\end{figure}

In addition to neutron-foil interactions, electrons can also be produced from material ablation at the foil. The optical image in figure~\ref{fig:foil_ablation}(a) shows bright conical structures around the blast shield and foil holder, indicating possible ablation. This could arise from foil heating due to x-ray or stray laser light, as suggested by the partially burnt foil in figure~\ref{fig:foil_ablation}(b). Further comparison between measured yield by the MRS and neutron activation detectors (NAD) in figure~\ref{fig:foil_ablation}(c) suggests an average material loss of $\sim 0.2 \ \mathrm{\mu m}$ for each implosion experiment, which is confirmed by pre- and post-exposure foil thickness measurements at General Atomics. However, this high-density region is not expected to play a role in energy modulations, as will be discussed at the end of \S\ref{sec:instability}. 

\begin{figure}
    \centering
    \includegraphics[scale=0.35]{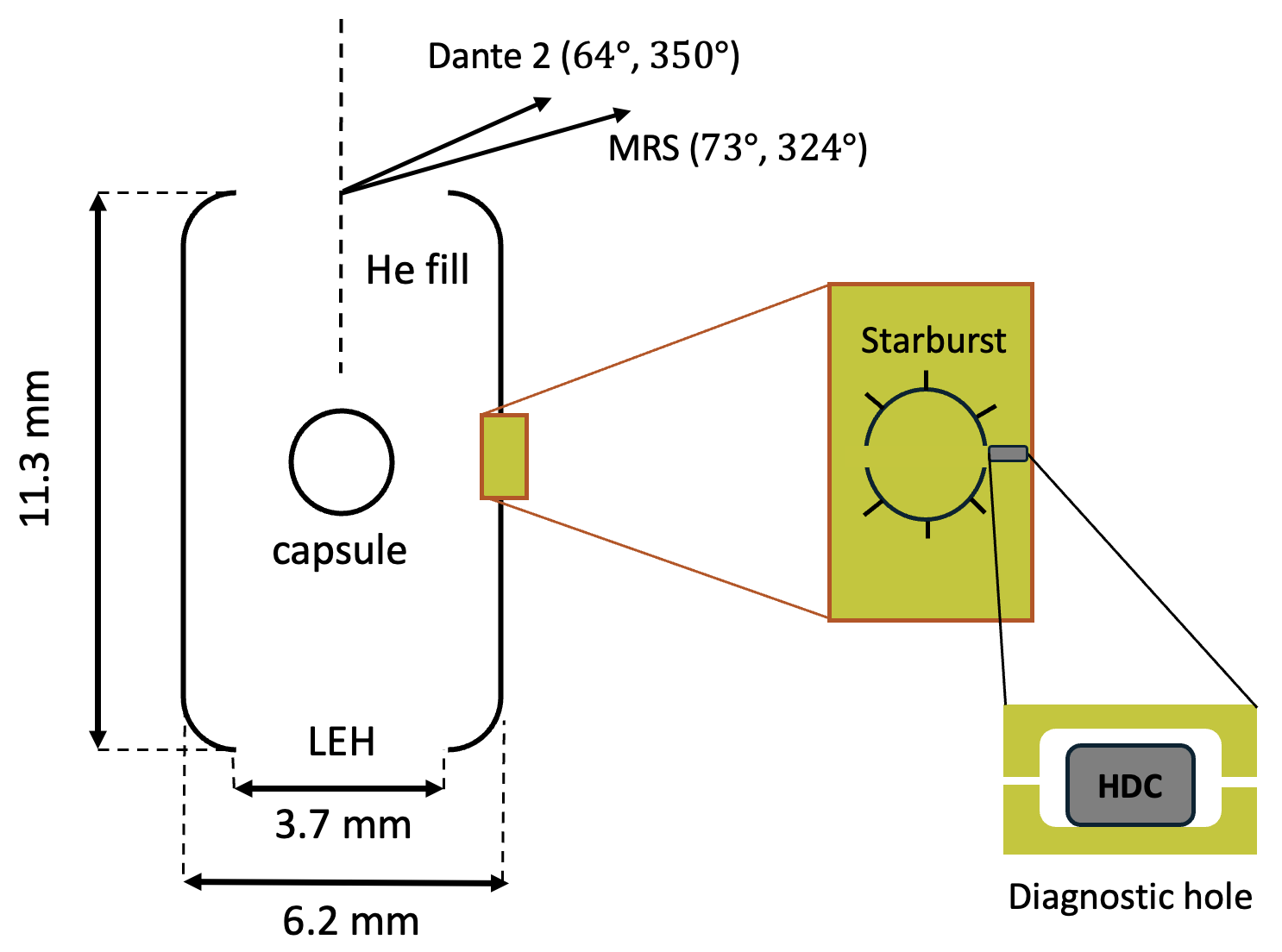}
    \caption{Schematic of the nominal NIF hohlraum design. Possible sources of x-ray leakage in the MRS direction include the starburst window, diagnostic hole (DH), and laser entrance holes (LEH). The DHs in other directions are not shown for clarity.}
  \label{fig:DH}
\end{figure}

Electrons can also be generated through x-ray photo-ionization of residual gas in the target chamber. Multiple x-ray sources contribute to this process. Inside the hohlraum, soft x-rays follow a Planckian distribution with a radiation temperature $T_r \sim 300$ eV. In addition, M-band radiation (photon energies $> 1.8$ keV) is emitted from the laser beam interacting with high Z coronal plasma wall blow-off \citep{lindl2004physics, dewald2020first}. Harder x-rays are generated from bremsstrahlung in the hotspot \citep{khan2018implementing} and from hot electrons interacting with the capsule target \citep{hohenberger2013measuring}. In this context, the x-ray flux along the MRS LOS located at spherical angles $(\theta, \varphi) = (73^{\circ}, 324^{\circ})$ needs to be determined. X-rays could escape in this direction through the openings annotated in figure ~\ref{fig:DH}. One such feature is the starburst window, an empty circular cut-out used to image the DT ice layer on the inner surface of the capsule ablator. This opening is predicted to close within $\sim 1-2$ ns after the laser drive begins \citep{milovich2010tuning}. Therefore, the x-ray leakage through the starburst is limited despite being directly in the MRS direction.

 Another leakage path is through the diagnostic hole (DH) located slightly off-axis to the MRS at $(90^{\circ}, \ 315^{\circ})$. HYDRA simulations \citep{marinak2001three} by \cite{milovich2020understanding} show x-ray leakage through the DH at early times via small gaps and at later times by burning through the high-density-carbon (HDC) slab shown in figure~\ref{fig:DH}. However, the DH contribution is expected to be minor since each DH subtends only $\sim~0.03$~sr from TCC. 

\begin{figure}
    \centering
    \includegraphics[scale=1]{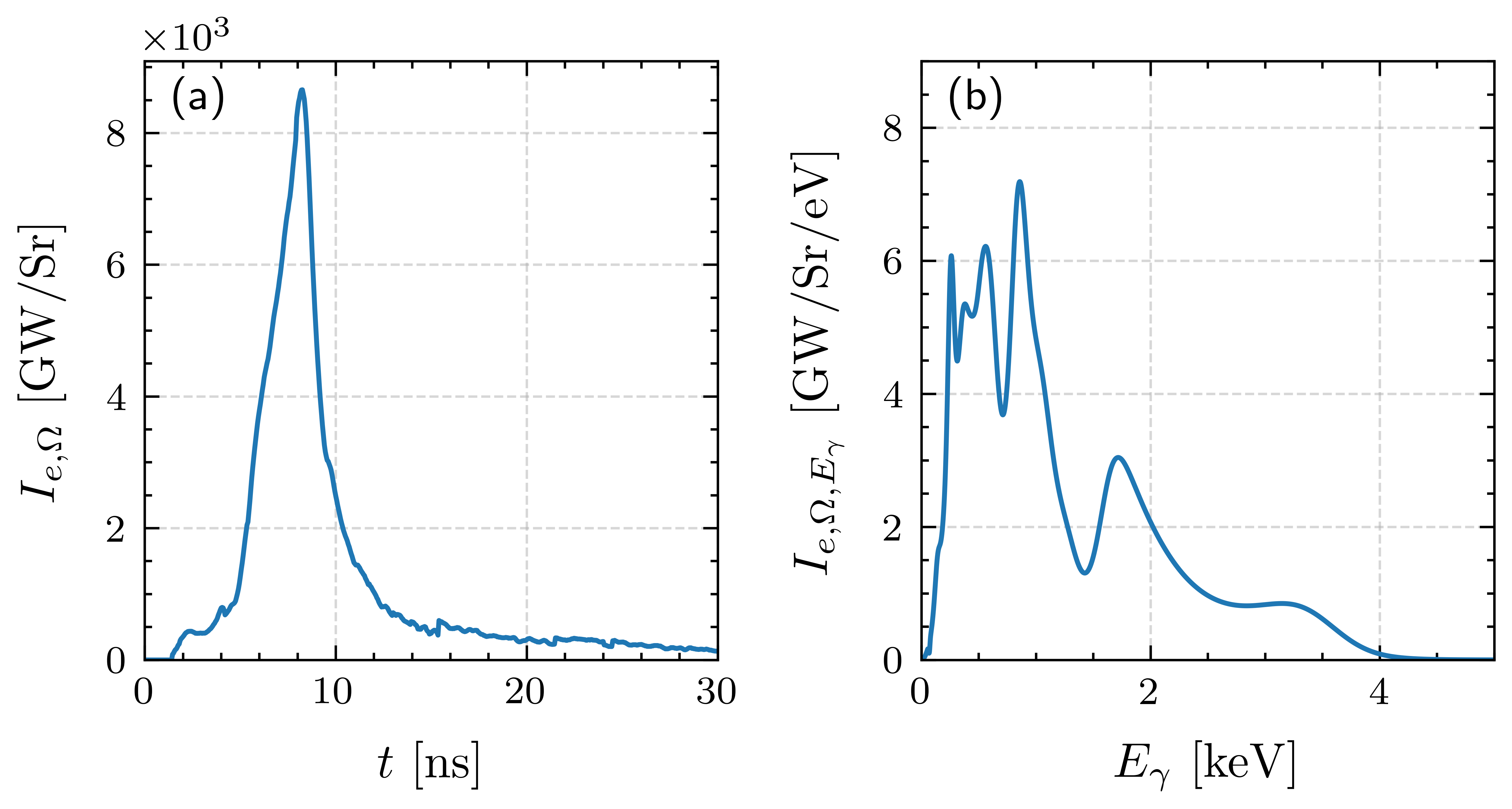}
   \caption[]{(a) Temporal profile of x-ray flux measured by Dante-2 for experiment N240504-001. (b) Measured x-ray energy spectrum at peak flux at $t\sim 8$ ns.}
  \label{fig:dante}
\end{figure}

The last and most probable source of leakage is through the laser entrance holes (LEH). The x-ray profile through the LEH is measured using Dante time-resolved x-ray spectrometers \citep{dewald2004dante}. Data from Dante-2 is considered due to its angular proximity to the MRS at $(64^{\circ}, \ 350^{\circ})$. The temporal profile of x-ray flux for experiment N240504-001 is shown in figure~\ref{fig:dante}(a), where the total radiant energy per unit solid angle is given by 
\begin{eqnarray}
    Q_{e, \Omega, D_2} = \int I_{e, \Omega} \ dt \sim 37 \ \mathrm{kJ \ sr^{-1}},
\end{eqnarray}
where $I_{e, \Omega}$ is the radiant intensity measured by Dante-2. Modeling the x-ray emission through the LEH as a Lambertian source \citep{decker1997hohlraum, may2015bright}, the total radiant energy per unit solid angle in the MRS direction is 

\begin{equation}
    Q_{e, \Omega, \text{MRS}} = \frac{Q_{e, \Omega, D_2} \cos(73^{\circ} ) }{\cos(64^{\circ})} \sim 25  \ \mathrm{kJ \ sr^{-1}}.
    \label{eq:flux_MRS}
\end{equation}

Equation~\eqref{eq:flux_MRS} neglects attenuation effects and therefore serves as an upper-bound estimate. Figure~\ref{fig:dante}(b) shows the x-ray energy spectrum at peak flux ($t\sim8$ ns), which resembles the expected Planckian profile on top of the M-band signature seen at $E_\gamma \sim 2 \ \text{keV}$.

Hard x-rays are typically measured using the FFLEX diagnostic along the equator \citep{hohenberger2013measuring}. Although FFLEX was not fielded on shot N240504-001, some simplified estimates can still be made. Soft x-rays are typically $\sim 1$~keV, while hard x-rays are typically $\gtrsim 10$~keV, resulting in an absorption factor $\sim 10^3$ higher for soft x-rays \citep{lemmon2010nist}. Moreover, the total energy of hard x-rays as measured by FFLEX is on the order of $\sim10$~kJ \citep{rinderknecht2015studies}, which is significantly less than the total energy of soft x-rays that escape through the LEH $\sim 100$~kJ \citep{chen2025drive}. For these reasons, soft x-rays emitted through the LEH are expected to be the dominant source of photo-ionization.

\subsection{Density and velocity distribution of photoelectrons}

The density of photoelectrons is estimated using the x-ray absorption cross-section $\sigma_{\text{vac}}$ of air from the NIST database \citep{lemmon2010nist}. Following the approach of \cite{swadling2017initial}, the x-ray attenuation due to residual gas along the MRS LOS is described by

\begin{eqnarray}
    I_{e,\Omega, \text{MRS}}(r) = \int I_{e,\Omega,\text{MRS}}(0) \exp\left(-\sigma_{\text{vac}}(E_\gamma) n_\text{vac} r \right)  dE_\gamma,
    \label{eq:attenuation_1}
\end{eqnarray}
where  $n_\text{vac}$ is the gas density inferred from the ambient pressure recorded for each experiment ($p_\text{vac} \sim 10^{-5}$ torr). The initial (unattenuated) radiant intensity $I_{e,\Omega, \text{MRS}}(0)$ seen by the MRS is split into a product of the source and spectral component 
\begin{eqnarray}
I_{e,\Omega,\text{MRS}}(0) = I_{e, \Omega, \text{MRS}} \cdot f(E_\gamma),
\label{eq:split}
\end{eqnarray}
where $f(E_\gamma)$ is the normalized photon spectrum approximated by the peak-flux profile in figure~\ref{fig:dante}(b). Substituting \eqref{eq:split} into \eqref{eq:attenuation_1} results in

\begin{eqnarray}
    I_{e,\Omega, \text{MRS}}(r) = I_{e,\Omega, \text{MRS}} \int f(E_\gamma) \exp\left(-\sigma_{\text{vac}}(E_\gamma) n_\text{vac} r \right)  dE_\gamma.
    \label{eq:attenuation_2}
\end{eqnarray}

The energy density deposited into the residual gas along the MRS LOS is then computed as 

\begin{equation}
    \frac{dE}{dV}(r)= \frac{d\Omega}{dA} \int \frac{-dI_{e,\Omega}(r) } {dr} \ dt.
\end{equation}

Taking the x-ray energy spectrum as approximately constant in time leads to the following expression

\begin{eqnarray}
\frac{dE}{dV}(r) = -\frac{Q_{e, \Omega,\text{MRS}}}{r^2} \frac{d}{dr} 
\int f(E_\gamma) \cdot \exp\left(-\sigma_{\text{vac}}(E_\gamma) n_\text{vac} r \right) 
\, dE_\gamma, 
\label{eq:ionization}
\end{eqnarray}

Assuming all deposited energy in \eqref{eq:ionization} is transferred to ionization, the maximum average charge state is obtained by solving

\begin{equation}
\frac{dE}{dV}(r)=n_\text{vac} \left[(\bar{Z}-\lfloor\bar{Z}\rfloor) I_{\lfloor\bar{Z}\rfloor-1}+\sum_{j=1}^{\lfloor\bar{Z}\rfloor} I_{j-1}\right],
\label{eq:state}
\end{equation}
where $I_j$ is the $j$ th ionization energy per atom and $\lfloor\bar{Z}\rfloor$ is $Z$ rounded down to an integer. This naturally leads to the electron density expressed as 

\begin{equation}
    n_e (r) = \bar{Z}(r)n_\text{vac}.
    \label{eq:ne}
\end{equation}

\begin{figure}
    \centering
    \includegraphics[scale=1]{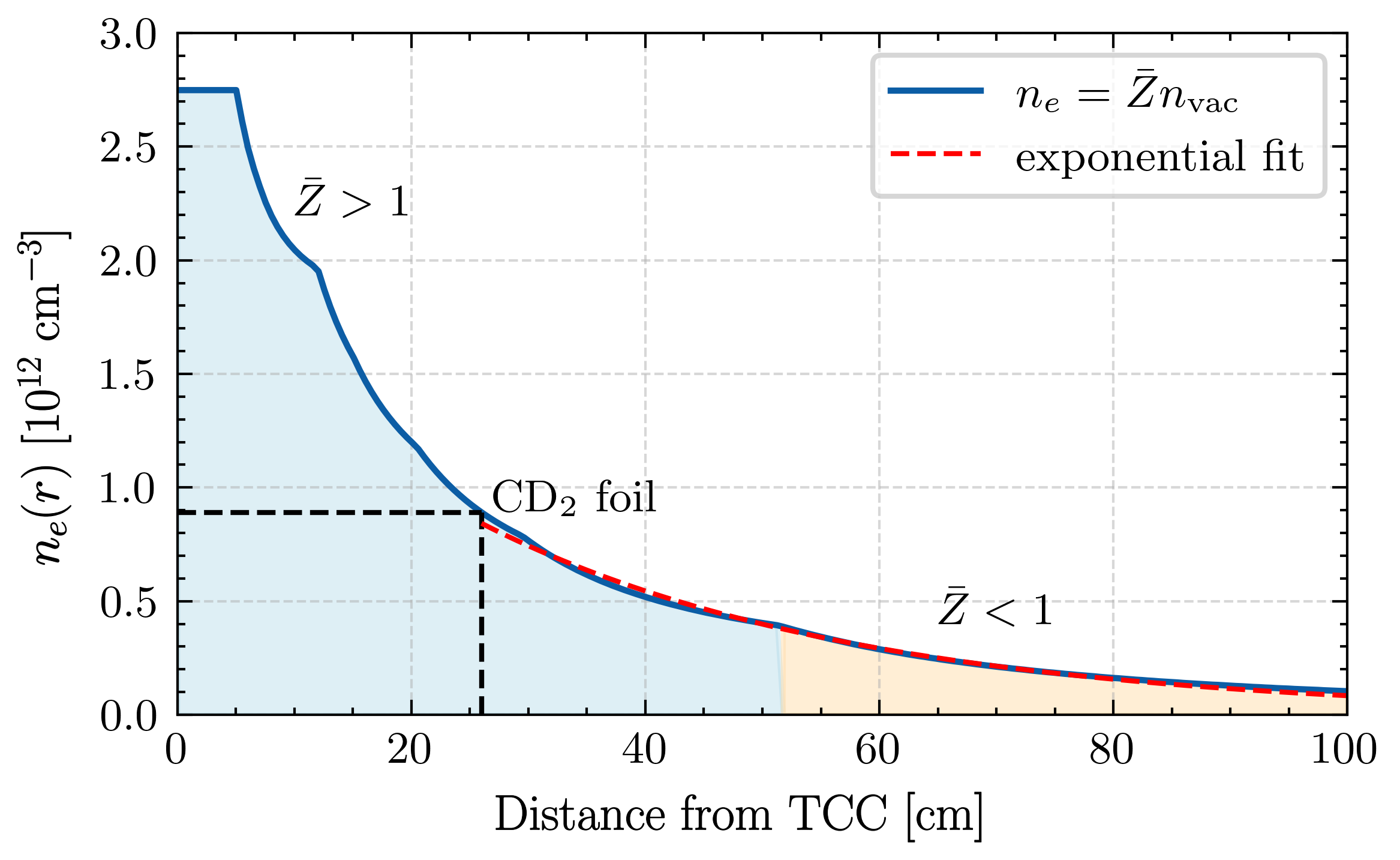}
   \caption[]{Estimated density profile of photo-ionized electrons. The blue region denotes $\bar{Z}>1$ while the orange region denotes $\bar{Z} < 1$. The foil and MRS aperture are located at $r=26$ cm and $r = 596$ cm, respectively. The initial flat profile indicates full ionization of nitrogen ($\bar{Z}=7$). }
  \label{fig:ionization_estimate}
\end{figure}

The numerical result of \eqref{eq:ne} is shown in figure~\ref{fig:ionization_estimate}, where the residual gas is assumed to be fully nitrogen for simplicity. Beyond the foil, the density of photoelectrons decays exponentially as given by

\begin{equation}
    n_e(r\geq 26  \ \text{cm}) = n_{e,f} \exp\left( \frac{-(r-26  \ \text{cm})}{L_n}\right),
    \label{eq:electron_dens}
\end{equation}
where $n_{e,f} \sim 9 \times 10^{11} \ \mathrm{cm^{-3}}$ and $L_n \sim 32$ cm are extracted from the exponential fit (red) in figure~\ref{fig:ionization_estimate}. These values serve as upper-bound estimates since they neglect attenuation and recombination effects.

Given their low density, photoelectrons cannot effectively thermalize within the experimental timescale. Their angular distribution must therefore be considered. For unpolarized light, this is described by

\begin{equation}
    \frac{dN_{e}}{d\theta_{\gamma,e}} \sim \sin\theta_{\gamma,e} - \frac{\beta(E_\gamma) \sin \theta_{\gamma,e}}{2}P_2(\cos\theta_{\gamma,e}),
    \label{eq:PI_angle}
\end{equation}
where $\theta_{\gamma,e} \in [0^\circ, 180^\circ]$ is the angle between the photon and electron velocity, $\beta$ is the asymmetry parameter, and $P_2$ is the second-order Legendre polynomial \citep{yeh1993atomic, reid2003photoelectron}. Since the x-ray energies are well above the ionization threshold ($\sim 10$ eV), the electrons follow the x-ray energy distribution biased by the absorption cross-section

\begin{eqnarray}
    \frac{dN_e}{dE_e} \sim f(E_\gamma) \cdot \sigma(E_\gamma).
    \label{eq:electron_energy}
\end{eqnarray}

\begin{figure}
    \centering
    \includegraphics[scale=0.4]{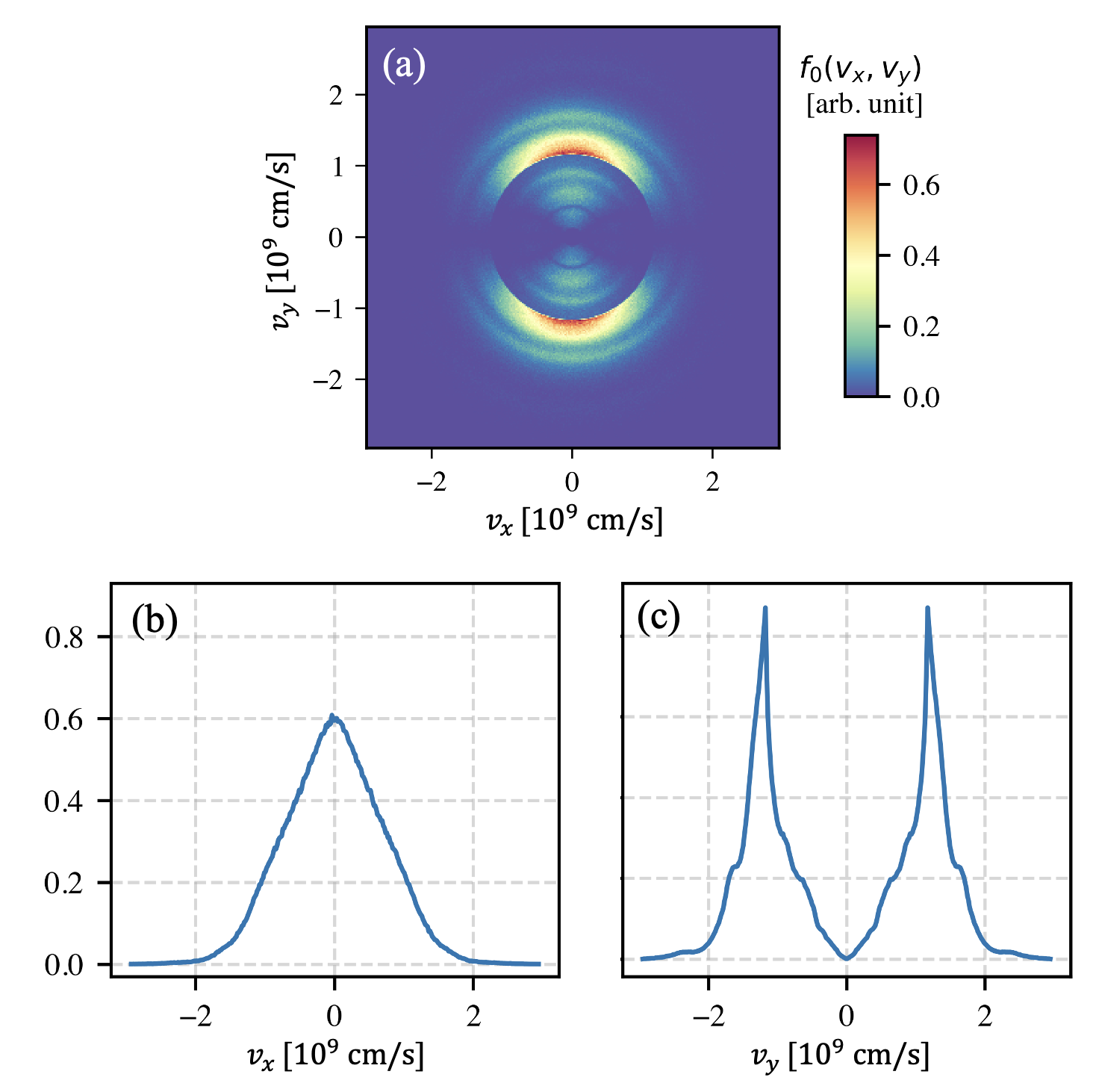}
   \caption[]{(a) 2D velocity distribution function of photo-ionized electrons, obtained by sampling the distribution in \eqref{eq:PI_angle} and~\eqref{eq:electron_energy}. Here, $v_{x}$ and $v_{y}$ correspond to the radial and angular components in the spherical coordinate system of the target chamber, respectively. (b) 1D velocity distribution in terms of $v_x$. (c) 1D velocity distribution in terms of $v_y$. }
  \label{fig:electron_vel}
\end{figure}

Combining the distributions in \eqref{eq:PI_angle} and~\eqref{eq:electron_energy} leads to the bimodal electron velocity distribution function (EVDF) in figure~\ref{fig:electron_vel}(a), where $v_{x}$ and $v_{y}$ correspond to the radial and angular components, respectively, in the spherical coordinate system of the target chamber. When the neutrons arrive at the foil, this EVDF is expected to have deviated from its initial profile due to electron–electron interactions, as will be discussed in \S\ref{sec:instability} and \S\ref{sec:pre-neutron}.

A natural question at this point is how the MRS blast shield factors into our previous estimates. The shield is made from a tantalum plate ($Z = 73$, radius $\sim$ 2 cm, thickness $\sim$ 1.57 mm) designed to block all debris and x-rays from the implosion. This creates a ``shadow region''  behind the shield where photo-ionization is suppressed. However, ionization can still occur in adjacent regions around the shield. Based on the estimated EVDF, electrons can travel 2 cm distance in $\sim$ 2 ns, giving them ample time to populate the shadow region before 14.1~MeV neutrons arrive at the foil ($\sim 5$~ns after the x-rays arrival at the foil radius).

\section{Analysis of beam-plasma instabilities}\label{sec:instability}

\subsection{Deuteron-electron modes}

To explore the possible ion–electron modes, a simplified model of deuterons streaming through a uniform plasma is considered. The deuterons are modeled as a beam with $\langle E_d \rangle \sim 12.5$ MeV, while the electrons follow the EVDF shown in figure~\ref{fig:flatten}(b). Collisional effects can be safely ignored due to the low electron density ($n_e \sim 10^{11} \ \mathrm{cm^{-3}}$). Under these conditions, the electrostatic or weakly electromagnetic instabilities are expected \citep{mima2018self}.

The electromagnetic beam-Weibel instability has a maximum growth rate of 

\begin{eqnarray}
   \gamma_\text{BWI} \sim \frac{\omega_b v_b}{c},
\end{eqnarray}
where $\omega_b = (n_de^2/m_d\epsilon_0)^{1/2}$ is the deuteron plasma frequency, $n_d$ is the initial deuteron density, $e$ is the elementary charge, $m_d$ is the deuteron mass, $\epsilon_0$ is the permittivity of free space, and $v_b$ is the beam velocity \citep{takabe2023theory}. Based on the reported yield, burn width, and foil setup, $n_d \sim 10^{10} \ \mathrm{cm^{-3}}$ near the foil rearside, which gives an e-folding time $\sim 70$ ns. Due to angular divergence of deuterons, the beam density drops significantly within this timescale, which renders this instability negligible. 

The electrostatic two-stream instability occurs when the beam velocity $(v_b \sim 3.4 \times10^9 \ \mathrm{cm / s})$ exceeds the electron velocity spread \citep{chen2012introduction}, a condition satisfied according to the EVDF illustrated in figure~\ref{fig:flatten}(b). The two-stream instability is evaluated by considering the following dispersion relation 

\begin{equation}
    D(\omega,k) = 1 + \sum_j \frac{\omega^2_{p,j}}{k_x} \int\frac{\partial \hat{f}_j (v_x) / \partial v_x}{\omega - k_xv_{x,j}} dv_x = 0,
    \label{eq:full_dispersion_relation}
\end{equation}
where $\hat{f}(v_x)$ is the normalized velocity distribution function, $k_x$ is the wave-vector component in the MRS direction, and $\omega_{p,j} = (n_j Z_j^2 e^2 / m_j \epsilon_0)^{1/2}$ is the plasma frequency \citep{chen2012introduction}. Here, $n_j$ is the density and $Z_j$ is the charge state of species $j\in \{\text{electron, ion, deuteron} \}$. For cold species, their velocity distributions are approximated as Dirac-deltas centered at the drift velocity $\hat{f}_j(v_x) = \delta(v_x - v_{d,j})$. Applying this to \eqref{eq:full_dispersion_relation} and integrating by parts gives 

\begin{equation}
    1 - \frac{\omega_\text{pe}^2}{\omega^2} - \frac{\omega_\text{pi}^2}{\omega^2} - \frac{\omega_b^2}{(\omega - k_x v_b)^2} = 0,
    \label{eq:dispersion_relation}
\end{equation}
where $\omega_\text{pe}$ and $\omega_\text{pi}$ are the electron and ion plasma frequencies of the ambient plasma. Since $\omega_\text{pi} \ll \omega_\text{pe}$, \eqref{eq:dispersion_relation} reduces to 

\begin{equation}
    (\omega^2 - \omega_\text{pe}^2)(\omega - k_x v_b)^2 = \omega_b^2 \omega^2.
\end{equation}

In the perturbative regime of $\alpha = n_b/n_e \ll 1$, the unstable mode occurs at $\omega\sim k_x v_b$. Assuming $\omega = k_xv_b + i\gamma$, the maximum TSI growth rate is 

\begin{equation}
    \gamma_\text{TSI} =  \frac{\sqrt{3} \, \omega_\text{pe}}{2} \left( \frac{m_e n_d}{2 m_d n_e} \right)^{1/3},
    \label{eq:growth_rate}
\end{equation}
which corresponds to an e-folding time of $\sim 3$ ns. Given that it takes $\sim$ 200 ns for aperture-selected deuterons to reach the MRS, there is more than enough time for TSI to develop. As the deuterons propagate, a longitudinal electric wakefield $\mathcal{E}_x$ is excited. For a thin beam with no angular variation \citep{hu2013modulation}, $\mathcal{E}_x$ is evaluated as 

\begin{equation}
\mathcal{E}_x(\xi) = \frac{-en_d}{\epsilon_0 k_p} \sin(k_p \xi) \mathrm{\Theta}(\xi),
\label{eq:E_field}
\end{equation}
where $\xi = x - v_bt$ is the longitudinal position in the beam's frame and $k_p=\omega_\text{pe}/v_b$ is the excited wavenumber. The Heaviside step function $\mathrm{\Theta(\xi)}$ imposes that the wake-field only forms behind the beam. This oscillatory field modulates deuteron energies through alternating regions of acceleration and deceleration. While the analytical approach provides a qualitative understanding, PIC simulations are required to self-consistently model the non-linear evolution of energy modulations.

\subsection{Electron-electron modes}

The bimodal EVDF represents two counter-streaming electron populations, which can drive the electron Weibel instability (EWI). The resulting magnetic field may deflect deuterons and therefore warrants further investigation. Adopting the analytical treatment of \cite{davidson1972nonlinear}, the bimodal EVDF shown in figure~\ref{fig:electron_vel}(a) is approximated by

\begin{eqnarray}
f_{0,e} (v_x,v_y) &\sim& \exp \left(-\frac{m_ev_x^2}{2k_bT_x} \right) 
\left[ \exp \left(-\frac{m_e(v_y - v_{d,e})^2}{2k_bT_y} \right) \right. \nonumber\\
&&\left. + \exp \left(-\frac{m_e(v_y + v_{d,e})^2}{2k_bT_y} \right) \right],
\label{eq:VDF}
\end{eqnarray}
where $k_b$ is the Boltzmann constant, $v_{d,e}$ is the electron drift velocity in the $y$ direction. The parameters $T_x \sim 290 \ \text{eV}$ and $T_y \sim 57 \ \text{eV}$ obtained from fitting the distributions in figure~\ref{fig:electron_vel}(b-c) characterize the directional velocity spread and do not imply thermal equilibrium. Here, the effective temperature in each direction is defined as

\begin{eqnarray}
    T_\parallel = T_x \label{eq:effective_temp_par} \\
    T_\perp = \frac{m_e}{2k_B} (\Delta v_{y} + v_{d,e})^2, 
    \label{eq:effective_temp_perp}
\end{eqnarray}
where $\Delta v_{y} = (2k_bT_y/m_e)^{1/2}$ is the velocity spread in the $y$ direction. The most unstable EWI occurs at

\begin{eqnarray}
    k_{\text{EWI}} = \frac{\omega_{pe}}{c} \left(\frac{T_{e \perp}}{T_{e \parallel}} - 1 \right)^{1/2},
\end{eqnarray}
with a corresponding growth rate given by

\begin{equation}
\gamma_{\text{EWI}} = \left( \frac{8}{27\pi} \right)^{1/2} \omega_{pe} \left( \frac{k_B T_{e\parallel}}{m_e c^2} \right)^{1/2} \frac{T_{e\parallel}}{T_{e\perp}} \left( \frac{T_{e\perp}}{T_{e\parallel}} - 1 \right)^{3/2}.
\label{eq:EWI_growth_rate}
\end{equation}

The corresponding e-folding time from \eqref{eq:EWI_growth_rate} is $\sim 10~\text{ns}$, allowing sufficient time for the EWI to develop. Over time, the fields saturate via magnetic trapping. Based on theory and 1D3V simulations \citep{davidson1972nonlinear}, the saturated magnetic field strength is given by

\begin{equation} |B_s| \sim \frac{\gamma_\text{EWI}^2  m_e c}{2e  k_\text{EWI} v_{d,e}}. \label{eq:B-field} \end{equation}

Evaluating \eqref{eq:B-field} gives \( |B_s| \sim 0.5 \ \text{mT} \). In the upper limit of a uniform field, 12.5 MeV deuterons are deflected by $\mbox{$\sim 0.3^\circ$}$, reducing the forward kinetic energy by $\mbox{$\sim 100$}$~keV. Since the electron density and Weibel-generated fields decay with distance from TCC, the actual downshift is likely less than $100$~keV. This is below the MRS energy resolution and thus considered negligible.


In addition to the EWI, oppositely streaming electrons can also drive the electron-electron TSI mode. The corresponding e-folding time, $\gamma_\text{TSI}^{-1} \sim 80 \ \text{ps}$, suggests that TSI can develop and saturate well before neutrons reach the foil. As kinetic energy is transferred into electrostatic waves, the initial bimodal EVDF described in \eqref{eq:VDF} is expected to flatten, consistent with previous work by \cite{hou2015linear}. This effect can be estimated using a simplified 2D PIC simulation, as will be discussed in \S\ref{sec:pre-neutron}.

\subsection{Foil-ablated electrons}

Up until now, our instability analysis has not considered the contribution of foil-ablated electrons. To estimate their impact, the upper limit in which the ablated plasma is fully ionized is considered. The inferred material loss per shot from figure~\ref{fig:foil_ablation}(c) is $\sim 10^{19}$ atoms, most of which is expected to occur during peak x-ray intensity. This time period can be approximated by the FWHM of the flux profile in figure~\ref{fig:dante}(a), $\tau_{\gamma} \sim 2.8 \ \text{ns}$. As an upper bound, the resultant plasma could reach a temperature $\sim 100$ eV and expand at the acoustic speed $c_s$, which gives a density scale length $L_{ab} \sim c_s \tau_\gamma \sim 100 \ \mu\text{m}$. The electron density can be described by the standard exponential profile

\begin{equation} 
n_\text{e, ab}(x) \sim n_0 \exp\left(\frac{-(x - 26 \ \text{cm})}{L_{ab}}\right), 
\label{eq:exp_profile} 
\end{equation}
where the peak electron density is $n_0 \sim 10^{21} \ \mathrm{cm^{-3}}$  \citep{atzeni2004physics}. 

The endpoint of \eqref{eq:exp_profile} should match the photo-ionized density $\sim 10^{11} \ \mathrm{cm^{-3}}$, which implies that the exponential profile spans a distance $\sim 0.2 \ \text{cm}$. Using $n_e = n_0$ in \eqref{eq:growth_rate}, the TSI growth time is $\gamma_\text{TSI}^{-1} \sim 40$ ps. This corresponds to a deuteron propagation distance of $\sim 0.14$~cm, which is barely within the valid range of \eqref{eq:exp_profile}. Energy modulation is a non-linear process that requires several growth times to develop. Therefore, foil-ablated electrons are excluded from subsequent simulations and analyses. Lastly, if the foil ablation is instead driven by stray laser light, the time scale $t_\text{laser} \sim 2.4$~ns can be used instead, which approximately corresponds to the peak duration of the laser drive. This would yield the same conclusion as for the x-rays case.

\section{Pre-neutron arrival simulation}\label{sec:pre-neutron}

Prior to the neutron arrival at the foil, photo-ionized electrons can interact through the EWI and TSI. This effect is simulated in 2D using the PIC code \textsc{epoch} \citep{arber2015contemporary, bennett2017users} which models the electron-electron interactions up to when neutrons arrive at the foil. The photo-ionized electrons are initialized with a uniform density $n_{e} = 10^{11} \ \mathrm{cm^{-3}}$ and velocities $(v_x, \ v_y)$ externally sampled from the bimodal profile in figure~\ref{fig:electron_vel}(a).  The simulation domain spans $\sim 10$ cm in each direction with a cell size $\Delta_x =\Delta_y = 0.5 \lambda_D$, where \(\lambda_D = \left( \epsilon_0 k_b T_e / m_e e^2 \right)^{1/2}\) is the Debye length. Here, $\lambda_D$ is calculated with $T_e~=~\text{min}(T_x, T_y)$ to ensure a sufficient spatial resolution. Periodic boundary conditions are used. Each cell consists of 128 fifth-order macro-particles, which helps suppress numerical heating and accurately resolve unstable modes \citep{mcmillan2020necessary}.

\begin{figure}
    \centering
    \includegraphics[scale=0.35]{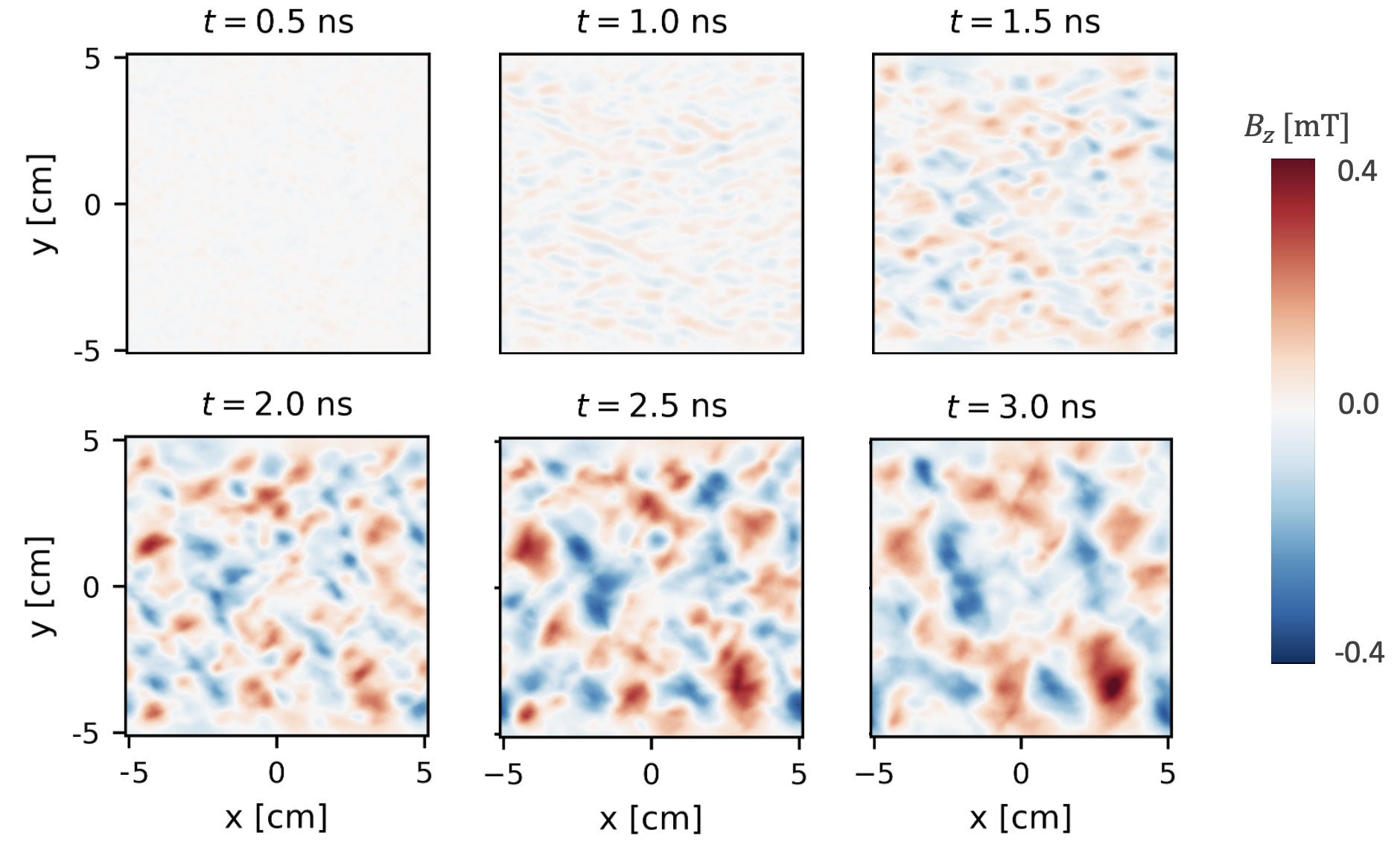}
   \caption[]{Temporal evolution of the magnetic field $B_z$ obtained from PIC. At $t=0$ (not shown), $B_z(x,y)=0$. The development of filaments as the magnetic field grows is characteristic of the electron-Weibel instability. The simulation box size is $\sim 5 \ \mathrm{cm}\times5 \ \mathrm{cm}$.}
  \label{fig:b-field}
\end{figure}

Figure~\ref{fig:b-field} shows the temporal evolution of the out-of-plane magnetic field $B_z$, where characteristic filamentation of the EWI is observed \citep{takabe2023theory}. The final magnetic field strength is $\sim 0.4$~mT, which is similar to the analytical estimate given in \S\ref{sec:instability}. As mentioned before, the magnetic field at this strength cannot meaningfully deflect deuterons. However, the filamentation in $B_z$ develops quicker than expected (predicted EWI growth time $\sim$ 10 ns). This is likely because the Gaussian approximations of \eqref{eq:VDF} does not account for the correlations between $v_x$ and $v_y$, which result in an underestimated anisotropy and growth rate. 

\begin{figure}
    \centering
    \includegraphics[scale=0.35]{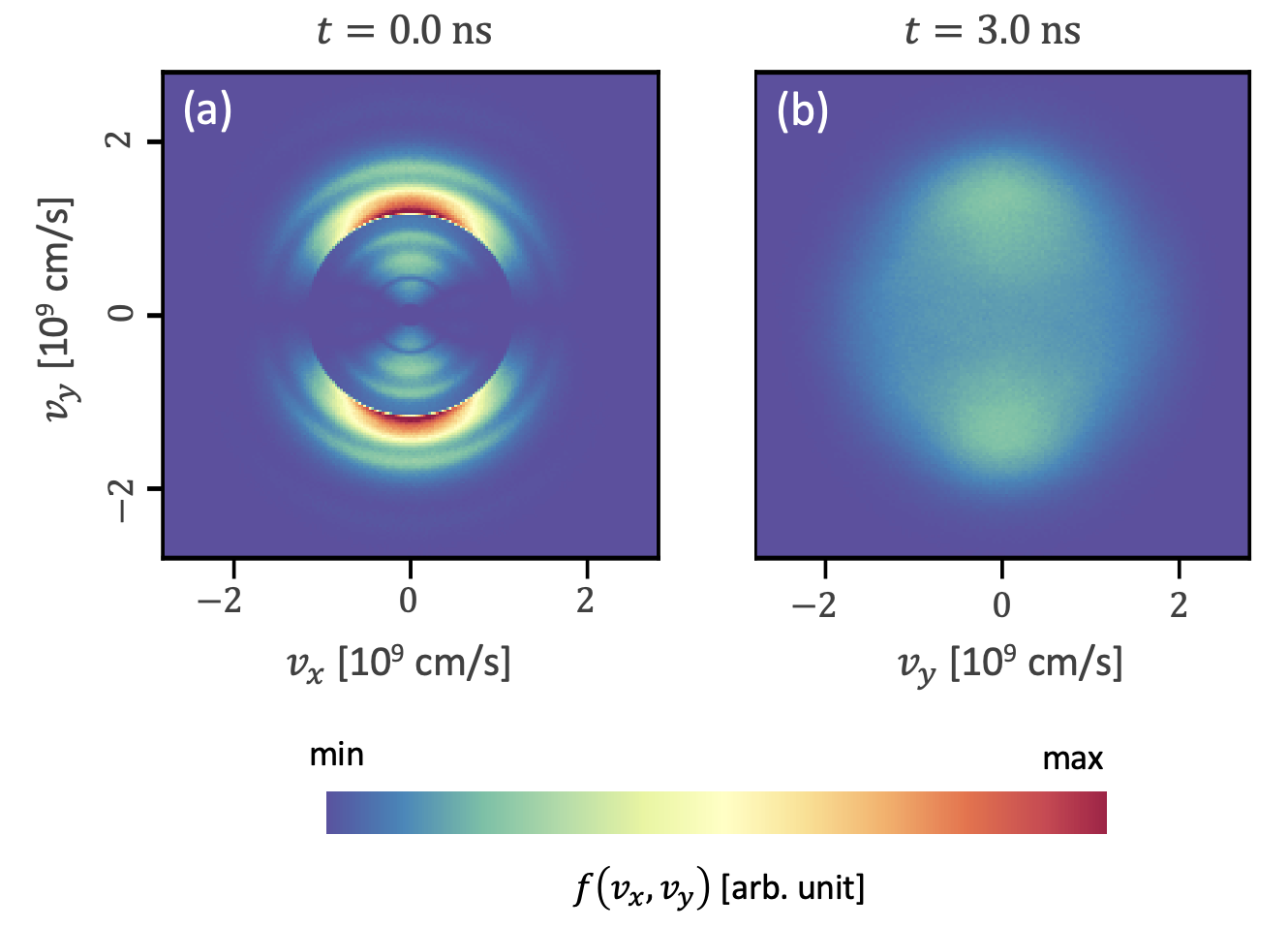}
   \caption[]{(a) Initial electron velocity distribution (bimodal). (b) Final electron velocity distribution (flattened).}
  \label{fig:flatten}
\end{figure}

The initial and final velocity distribution of electrons as determined from the 2D PIC simulation are shown in figure~\ref{fig:flatten}(a-b). As expected from TSI interactions, the initial bimodal structure in $v_y$ becomes merged and flattened \citep{hou2015linear}. The final distribution in figure~\ref{fig:flatten}(b) then serves as input for the post-neutron arrival (at the foil) simulations.

\section{Post-neutron arrival simulation}\label{sec:post-neutron}
\subsection{Simulation set-up}\label{subsec:sim_set_up}

\begin{figure}
    \centering
    \includegraphics[scale=0.35]{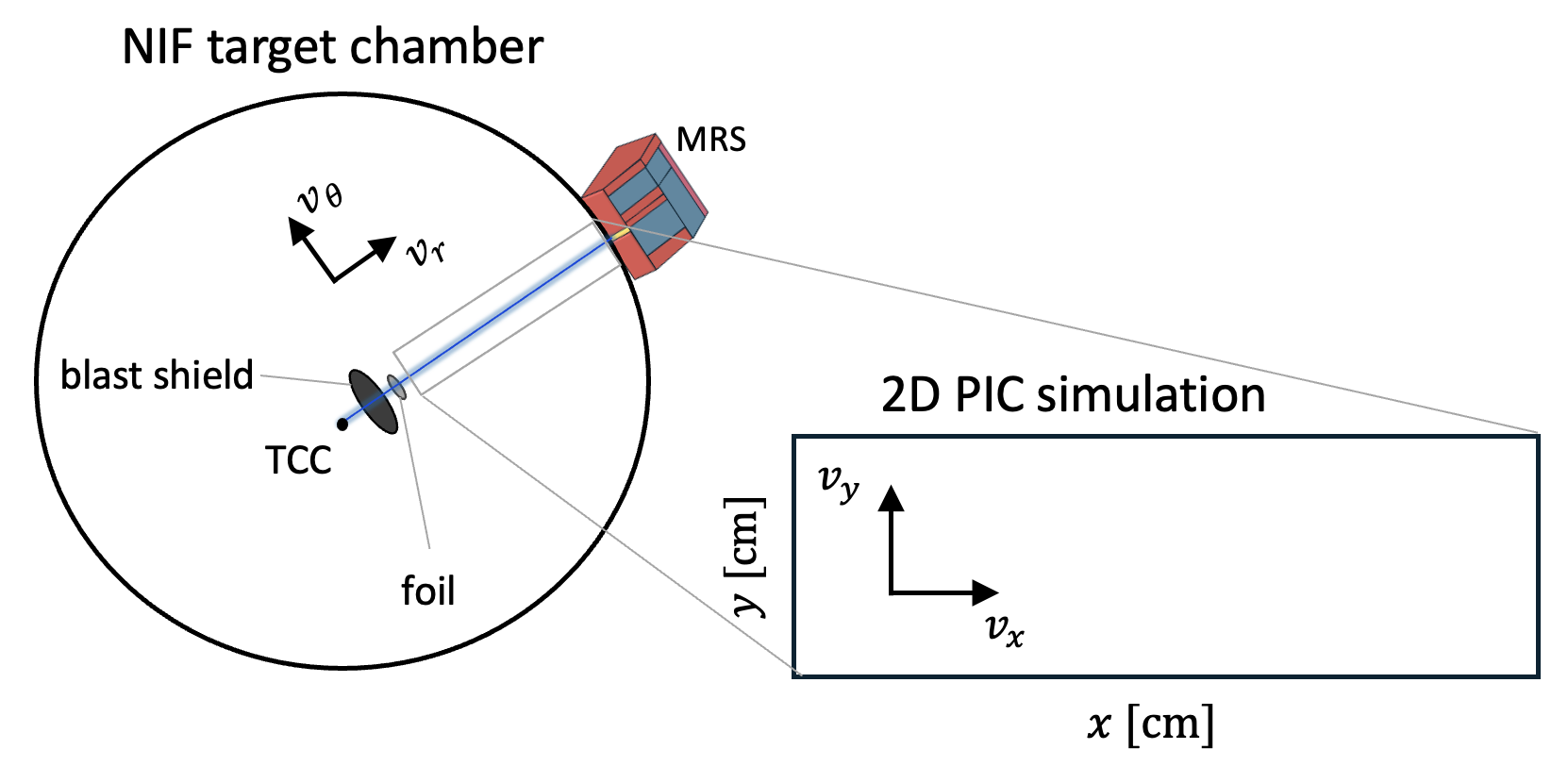}
   \caption[]{Diagram of the 2D PIC simulation frame relative to the spherical target chamber geometry (not to scale). The blue line shows the MRS line of sight.}
  \label{fig:PIC_set_up}
\end{figure}

The post-neutron arrival simulations model the transport of deuterons through a tenuous plasma from the foil backside to the MRS magnet aperture (see figure~\ref{fig:PIC_set_up}). The initial domain spans $x \in [26, 150]$~cm and $y \in [-1, 1]$~cm with a uniform grid size $\Delta_x  = \Delta_y = 0.7\lambda_D$. A moving window is applied to ensure aperture-selected deuterons can reach $x = 596$~cm, where a particle probe representative of the MRS records their kinetic energies for direct comparison with the experiment.

The ambient electrons are initialized with 150 fifth-order macro-particles per cell based on the density profile in \eqref{eq:electron_dens} and modified EVDF described in \S\ref{sec:pre-neutron}. Due to uncertainties in the x-ray flux, the term $n_{e,f}$ and $L_n$ are treated as free parameters. A reasonable match with the experiment is obtained for $n_{e,f} \sim 2$–$3 \times 10^{11} \ \mathrm{cm^{-3}}$ and $L_n \sim 15$–20 cm. Compared to the estimates in \S\ref{sec:exp_conditions}, these values are slightly lower by a factor of $\sim3$-$5$ times but remain within the same order of magnitude.

\begin{figure}
    \centering
    \includegraphics[scale=0.31]{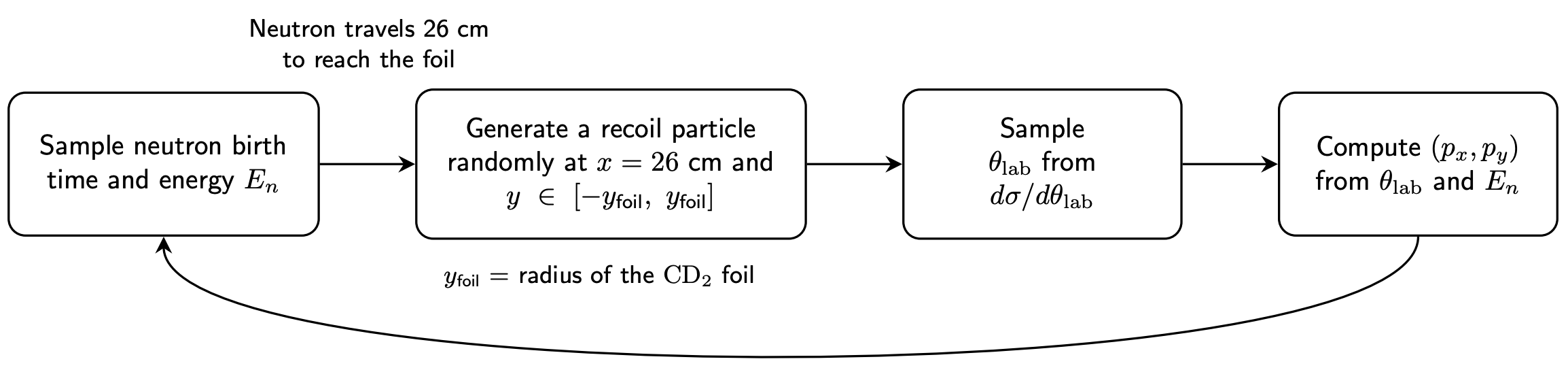}
   \caption[]{External sampling routine of recoil particles. The neutron energy is sampled based on the reported $(Y_n, \rho R, T_i)$, accounting for Doppler-broadening. Its birth time is drawn from a Gaussian burn duration with a FWHM of $\tau_\text{BW}$. A recoil particle is generated once the neutron hits the foil. These steps are repeated until sufficient sampling is reached.}
  \label{fig:sampling}
\end{figure}

Recoil protons and deuterons are each initialized with $2~\times~10^8$ macro-particles. Their weight is determined from the particle numbers in Table~\ref{tab:geant_particle_output}, scaled according to the reported neutron yield and foil configuration of experiment N240504-001. Due to the complex initial phase-space distributions, the recoil particles are externally sampled and loaded into the simulation via binary files (see figure ~\ref{fig:sampling}). The built-in particle injector in \textsc{epoch} is avoided to prevent the formation of a large sheath electric field, which could distort the initial deuteron energy spectrum.

\subsection{Simulation results}\label{subsec:sim_results}

\begin{figure}
    \centering
    \includegraphics[scale=0.35]{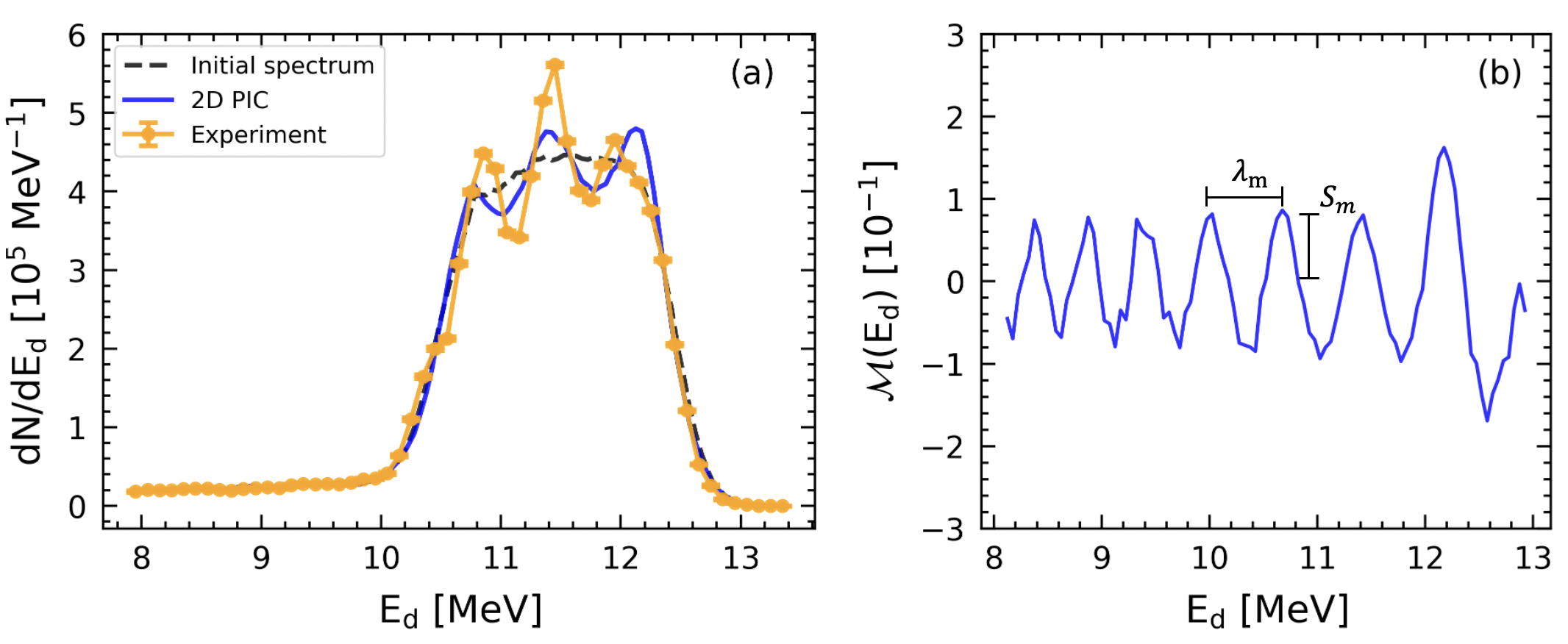}
   \caption[]{(a) Comparison between the simulated (blue) and experimental (orange) deuterons energy spectrum. The dashed black line shows the assumed initial spectrum. Note that the  error bars are smaller than the plot markers. (b) Normalized modulation profile of the simulated spectrum calculated from \eqref{eq:mod_norm}.}
  \label{fig:spectrum}
\end{figure}

A non-exhaustive scan indicates that $n_{e,f}\sim2.4\times10^{17} \ \mathrm{m^{-3}}$ and $L_n \sim 15$~cm provide a reasonable match to the experiment. The result is shown in figure~\ref{fig:spectrum}(a), where the discrepancy between the the simulated and measured spectra could be due to missing details in the electron distribution.

The energy modulation is directly quantified in figure~\ref{fig:spectrum}(b). Given the initial spectrum $f_0$ and modulated spectrum $f_1$, the normalized modulation profile is defined as

\begin{equation}
    \mathcal{M}(E_d) = \frac{f_1(E_d) - f_0(E_d)}{f_0(E_d)}.
    \label{eq:mod_norm}
\end{equation}

\begin{figure}
    \centering
    \includegraphics[scale=0.34]{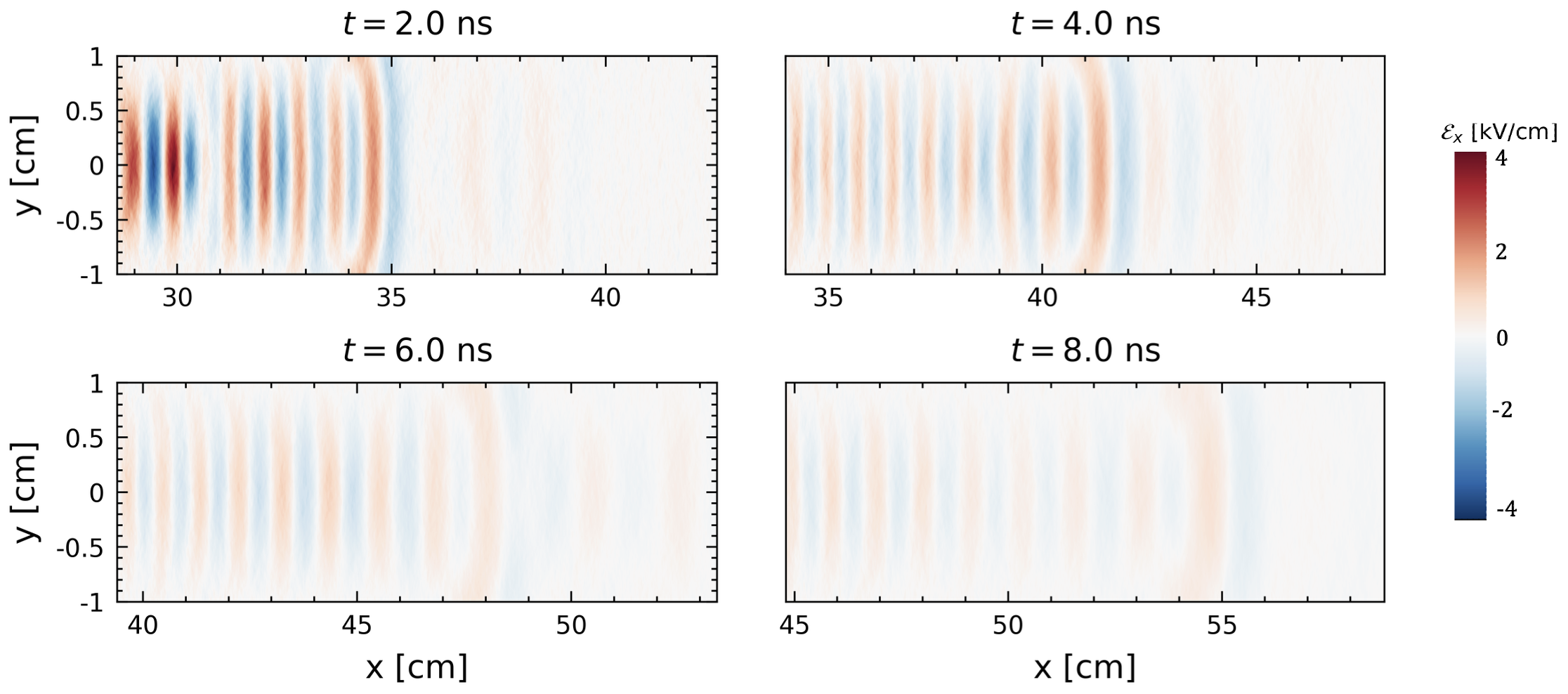}
   \caption[]{Zoomed-in view of the longitudinal electric field at various times. After 8 ns (not shown), the sinusoidal structures still persist but with reduced amplitudes, making them less visible with the given color scale. The x-axis values vary due to the applied moving window.}
  \label{fig:e-field}
\end{figure}

The simulated $\mathcal{M}(E_d)$ profile in figure~\ref{fig:spectrum}(b) shows an approximately uniform modulation wavelength $\lambda_m \sim 600 \ \text{eV}$ and modulation amplitude $S_m \sim 0.1$. To understand these results, the beam-plasma instability analyses  in \S\ref{sec:instability} are revisited. By comparing the ratio of the magnetic energy to electric energy at various times

\begin{eqnarray}
    \frac{\int |B|^2 \ dV}{\int \mu_0\epsilon_0 |\mathcal{E}|^2 \ dV} \lesssim 10^{-2},
    \label{eq:energy_ratio}
\end{eqnarray}
it is deduced that the contribution from the beam-Weibel instability is relatively minor. Figure~\ref{fig:e-field} shows spatially oscillatory structures in $\mathcal{E}_x$ at various times, which is indicative of the electrostatic two-stream instability. The observed spatial wavelength at $t=2$~ns, for example, is $\mbox{$\sim0.8$}$~cm, which agrees with the analytical prediction $\lambda_\text{TSI} = 2\pi v_b / \omega_\text{pe}$ at the local plasma frequency. In addition, the induced electric field is kV/cm in magnitude, as predicted by~\eqref{eq:E_field}. The induced electric field dimnishes over time due to angular divergence of the deuterons.

\begin{figure}
    \centering
    \includegraphics[scale=0.4]{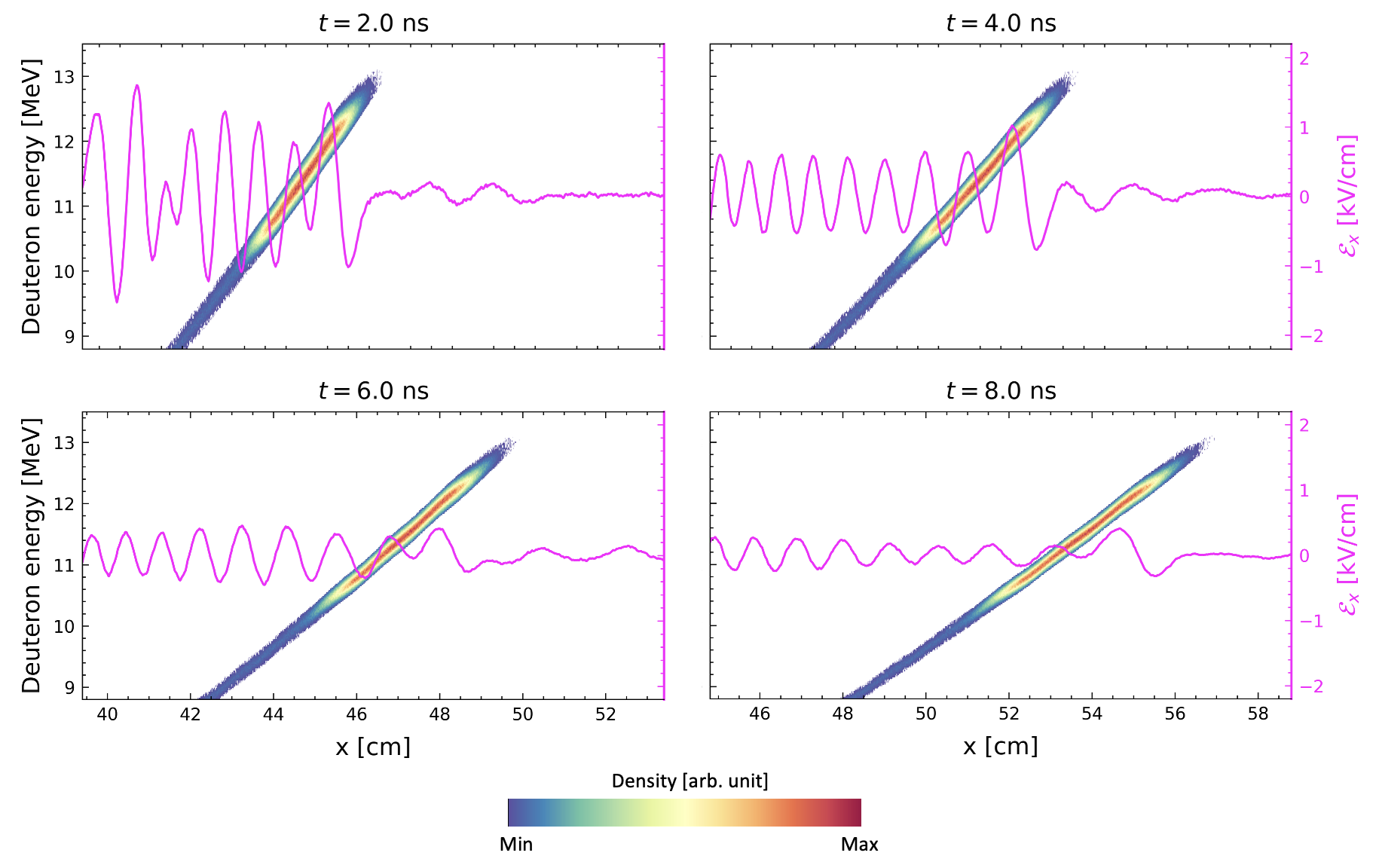}
   \caption[]{Energy phase space diagram of deuterons at the same times as shown in figure~\ref{fig:e-field}. The longitudinal electric field averaged along the y-axis is overlaid in pink.} 
  \label{fig:phase_space}
\end{figure}

\begin{figure}
    \centering
    \includegraphics[scale=0.42]{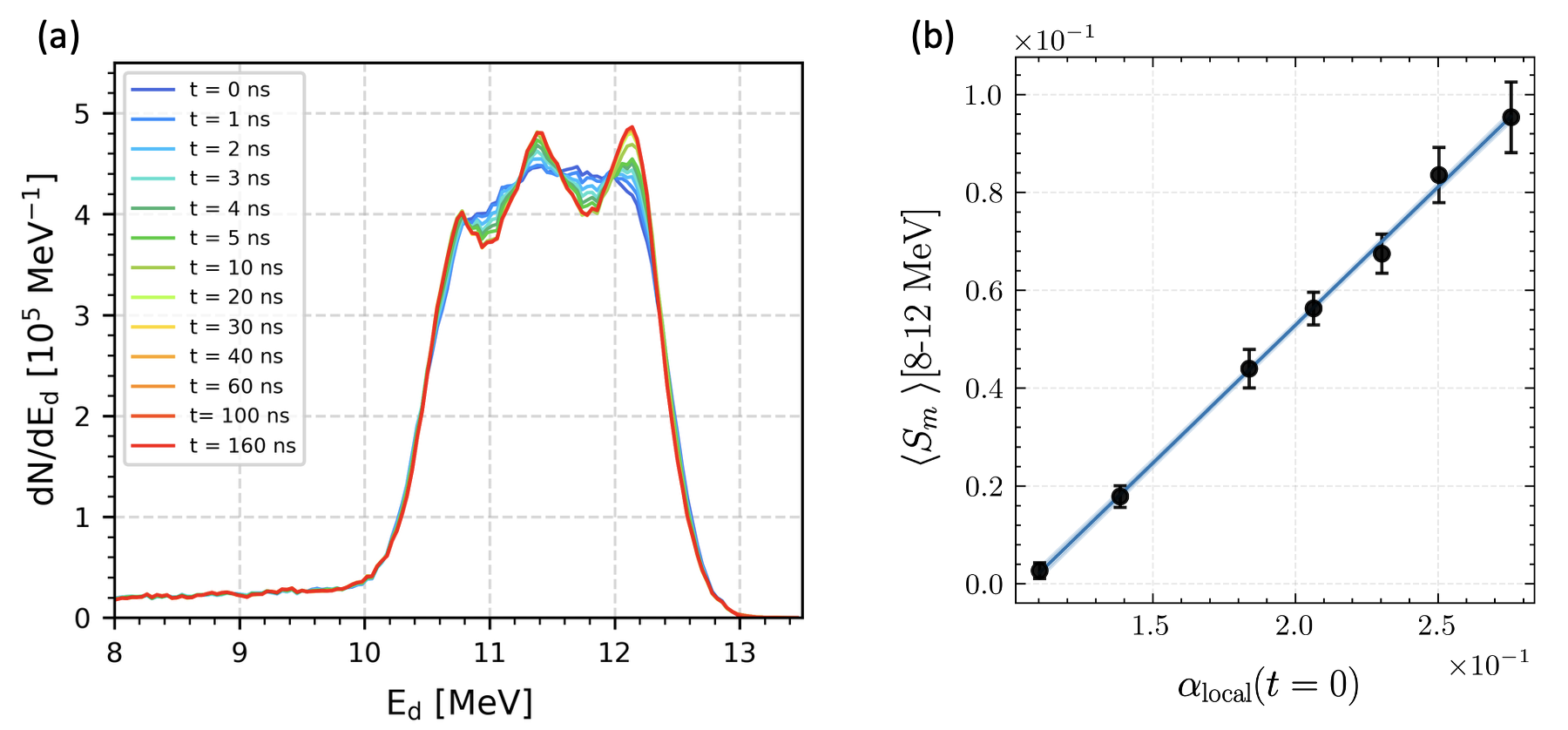}
    \caption[]{(a) Deuteron energy spectra at selected times. Modulations mostly occur in the early times when the induced electric field is stronger. Saturation is observed at around $t \sim 60$~ns. (b) Modulation amplitude averaged between 8-12 MeV as a function of the local density ratio at $t = 0$. The vertical error bars represent the standard deviation of each data point, while the light blue band indicates the standard deviation of the fit.}
    \label{fig:scan}
\end{figure}

To make the connection between spatial and energy modulations, we analyze the energy phase space diagram in figure~\ref{fig:phase_space}, where the excited $\mathcal{E}_x$ averaged along the y-axis is overlaid in pink. At a given time $t=t_1$, deuteron energy relates to the longitudinal position through

\begin{equation}
E_d(x, t_1) = a(t_1) x + b(t_1),
\end{equation}
where $a(t_1)$ and $b(t_1)$ parameterize the straight line in the energy phase space diagram. In the simulation, the wavelength of energy modulation stabilizes by $t \sim 2$ ns, justifying the constant $a(t_1 = 2 \ \text{ns})$ for our purposes. Assuming this behavior is similar across different $(n_d, n_e)$, it is expected that $\mbox{$\lambda_\text{mod} \sim a(t_1 = 2 \ \text{ns}) \lambda_\text{TSI}$}$, which scales with $n_e^{-1/2}$. The other modulation quantity of interest is the amplitude $S_\text{m}$, which is expected to scale with the amplitude of the electrostatic potential $|\phi(x)|$ based on dimensional considerations. Integrating \eqref{eq:E_field} leads to

\begin{equation}
    \phi(\xi) = \frac{en_d}{\epsilon k_p^2} \cos(k_p \xi) \Theta(\xi), 
    \label{eq:potential}
\end{equation}
which scales proportionally with $\alpha = n_d/n_e$. To validate these scaling relations and gain a broader understanding of modulations, $(n_d, n_e)$ are systematically varied based on the realistic yield, laser energy and vacuum pressure. A full-scale simulation for each set of $(n_d,n_e)$ would be computationally extensive. Therefore, the simulation run time is truncated at 60 ns, where saturation in energy modulations is observed in figure~\ref{fig:scan}(a). 

To characterize the modulation amplitude, $\langle S_m \rangle$ is defined as the average peak height within the range $E_d \in [8,12]$MeV, where the amplitude remains approximately uniform. As shown in figure~\ref{fig:scan}(b), $\langle S_m \rangle$ increases linearly with the initial local density ratio $\alpha(t=0)$, consistent with the scaling from \eqref{eq:potential}. In contrast, the modulation $\lambda_m$ is observed to decrease as $n_e$ increases. Within the realistic $n_e$ range, $\lambda_m$ varies between 500 to 700~keV. However, the difference is small and not always detectable within the energy bin width of 50 keV. There are two important caveats about these results. First, this scaling is valid only in the perturbative regime. For $\alpha \gtrsim 1$, the electric field evolution can become highly non-linear due to the formation of forerunner electrons \citep{hara2018generation}, which can significantly alter the deuteron energy spectrum. This effect arises in the non-linear phase of TSI when a fraction of the initially trapped electrons become de-trapped and accelerate ahead of the deuteron beam, which drives a secondary TSI with the background electron population. Second, sufficient ionization is required to sustain a plasma. At very low electron densities, TSI is expected to vanish, and interactions with neutrals are expected to dominate.

\section{Sensitivity study}\label{sec:sensitivity_study}

A synthetic-data study was conducted to assess the impact of modulations on high-yield experiments. The synthetic neutron spectrum is initialized by parameters $(Y_n$, $T_i$, $\rho R)$. The primary component is modeled as a Gaussian distribution with its mean and width determined by the ion temperature $T_i$, while the down-scatter component scales linearly with $\rho R$ \citep{ballabio1998relativistic}. The model neutron spectrum is then convolved with the MRS response function to give the measured deuteron spectrum $f_0(E_d)$. To mock-up energy modulations in the deuteron spectrum, a sinusoidal profile with uniform amplitude and wavelength is used

\begin{equation}
    \mathcal{M} (E_d)= 1 + S_\text{m} \sin \left(\frac{2\pi}{\lambda_\text{m}} (E_d + \delta E) \right),
\end{equation}
where  $\delta E \in [-50,50]$ keV is a random energy shift. The resultant modulated spectrum is expressed as 

\begin{equation}
    f_\text{1}(E_d) = f_0(E_d)\mathcal{M}(E_d) \cdot \frac{N_d}{N_\text{m}},
\end{equation}
where the term $N_d/N_\text{m}$ redistributes the deuterons to conserve the total count. 

Afterwards, Poisson-distributed noise and realistic background subtraction are applied to each energy bin of $f_1(E_d)$. The background accounts for intrinsic plastic defects and neutron-induced tracks in CR-39 analysis, which can affect the measured deuteron spectrum in the down-scattered region \citep{casey2011coincidence}. With the modulated deuteron spectrum generated, neutron yield, $T_i$, and $\rho R$ are determined from the optimal fit parameters. Our fitting routine employs the \texttt{curve\_fit} function from SciPy library \citep{virtanen2020scipy}, where more weight is given to data points with lower uncertainties. The statistical uncertainty in each energy bin is estimated based on the deuteron count and background contribution.  

Analytical theory and simulation results suggest the main levers of energy modulations are the deuteron $n_d$ and background electron density $n_e$. The deuteron density scales as 

\begin{equation}
    n_d = \frac{N_d}{V_d} \propto \frac{Y_n d_f A_f}{\tau_\text{BW}}
    \label{eq:nd_scale}
\end{equation}
where $V_d$ is the initial volume occupied by deuterons, $d_f$ is the foil thickness, and $A_f$ is the foil area. Since photoelectrons are produced by soft x-rays ionizing the residual gas in the target chamber, $n_e$ is expected to scale with the laser drive energy $E_L$ and residual gas pressure $p_\text{vac}$

\begin{equation}
    n_e \propto E_L p_\text{vac}.
    \label{eq:ne_scale}
\end{equation}

Combining \eqref{eq:nd_scale} and \eqref{eq:ne_scale} yields the scaling for $S_m$

\begin{equation}
    S_m \sim \frac{n_d}{n_e} \propto \frac{Y_n d_f A_f}{E_L p_\text{vac} \tau_\text{BW}},
    \label{eq:S_scale}
\end{equation}
where the modulation amplitude $S_m$ correlates with yield. In contrast, the modulation wavelength $\lambda_m$ is less restricted and therefore allowed to vary between 500 to 700~keV. To simplify the analysis of the synthetic study, the burn width $\tau_\text{BW}$ and vacuum pressure $p_\text{vac}$ are assumed as fixed. This leaves $(Y_n, d_f, A_f, E_L)$ as the driving factors for modulations. For the scaling in \eqref{eq:S_scale}, shot N240504-001 is used as the reference point.

\begin{figure}
    \centering
    \includegraphics[scale=0.48]{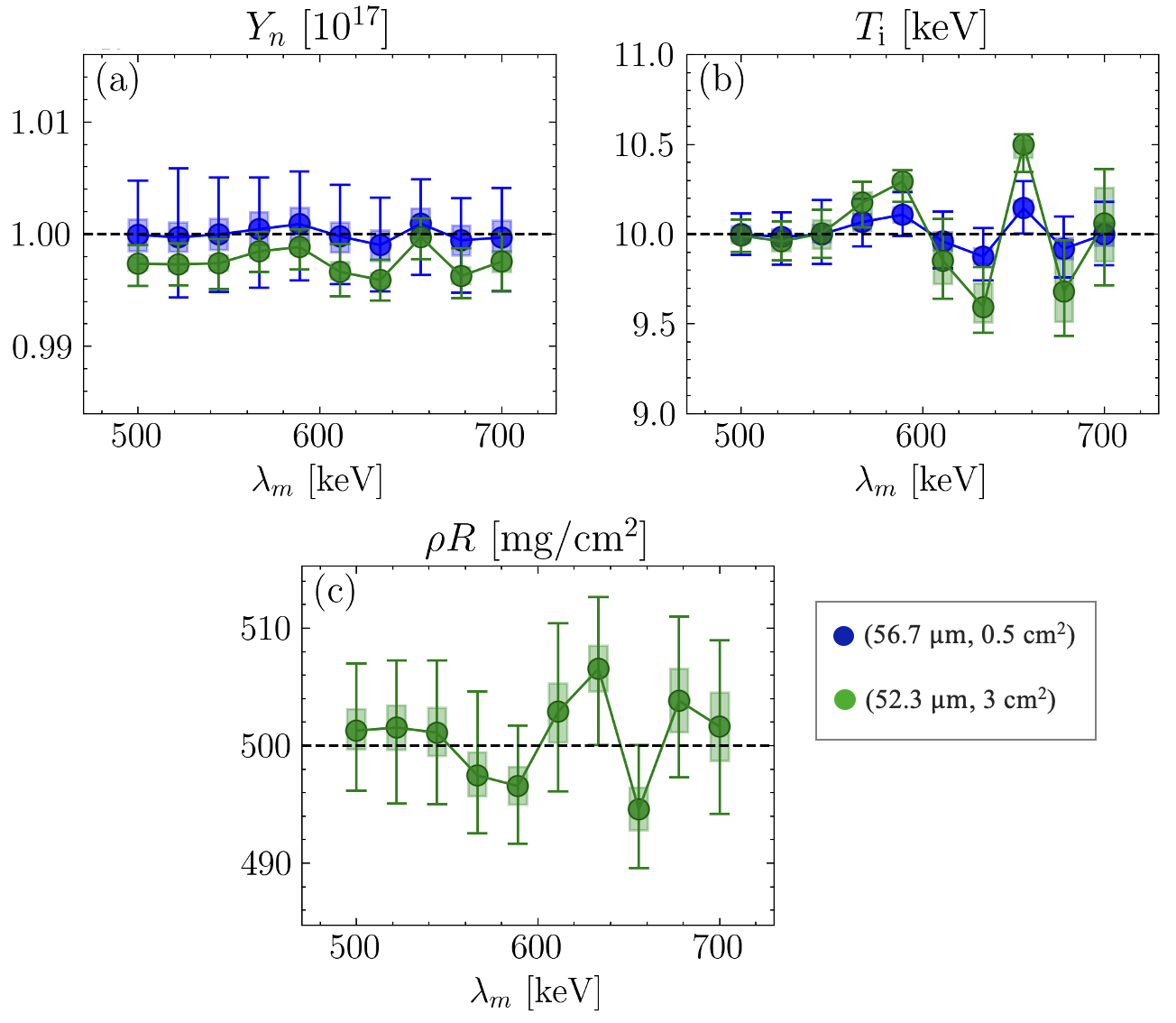}
    \caption[]{Effect of energy modulations on the inferred (a) neutron yield, (b) ion temperature, (c) areal density. The color shading represents the standard deviation while the error bars show the maximum and minimum values over 1000 synthetic spectra. Here, $Y_n \sim 10^{17}$, $\rho R \sim 500 \ \mathrm{mg/cm^2}$, $T_i \sim 10$ keV, and $E_L \sim 2$~MJ. Note that in practice the $(56.7 \ \mathrm{\mu m}, 0.5 \ \mathrm{cm^2})$ foil configuration is not used to infer $\rho R$. }
    \label{fig:synthetic}
\end{figure}

A typical high-yield implosion could have the following parameters: $Y_n \sim 10^{17}$, $T_i \sim 10$ keV, $\rho R \sim 500 \ \mathrm{mg/cm^2}$ and $E_L \sim 2$~MJ. The foil configurations $(d_f, \ A_f)$ in these experiments include $(52.3 \ \mathrm{\mu m}, \ 3 \ \mathrm{cm^2})$ and $(56.7 \ \mathrm{\mu m}, \ 0.5 \ \mathrm{cm^2})$. Figure~\ref{fig:synthetic} shows the effect of energy modulations on the MRS-inferred parameters averaged over 1000 synthetic spectra. For both foil configurations, the results are are within the expected systematic uncertainties: $\Delta Y_n/Y_n \leq 5.5\%$, $\Delta \rho R/ \rho R \sim 5\%$, and $\Delta T_i\sim 0.46$ keV. Therefore, the impact of modulations remains negligible at $Y_n \sim 10^{17}$, which also explains why they have only recently become apparent in the data.

\begin{figure}
    \centering
    \includegraphics[scale=0.5]{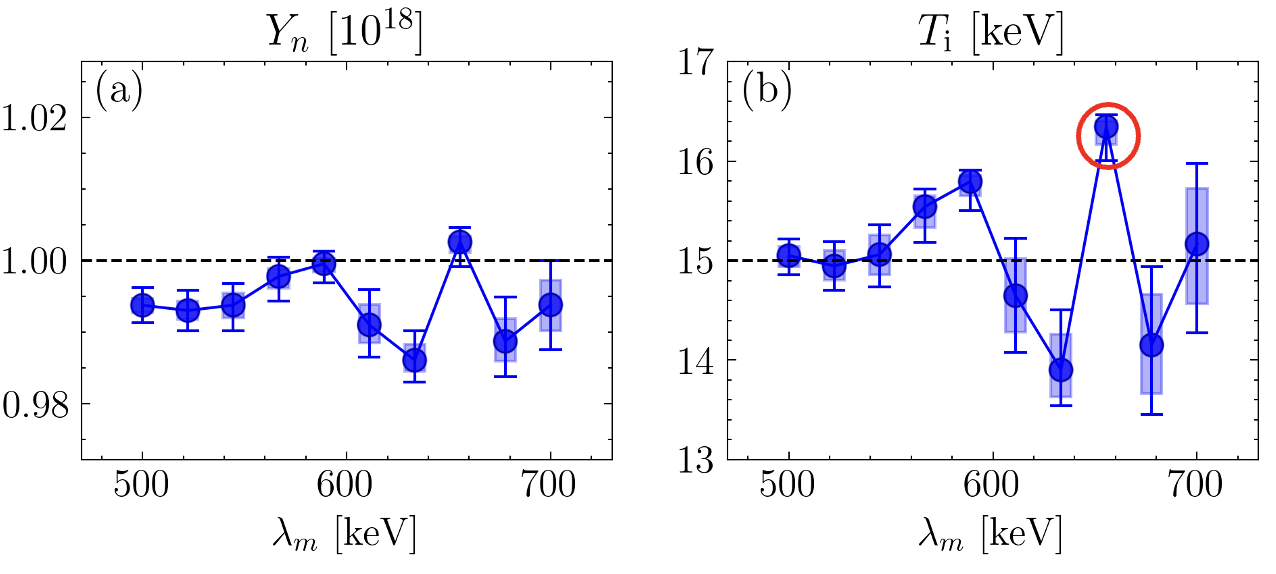}
    \caption[]{Effect of energy modulations on the inferred (a) neutron yield, (b) ion temperature of the $(56.7 \ \mathrm{\mu m}, 0.5 \ \mathrm{cm^2})$ foil configuration. The red circle indicates the data point outside the acceptable range. The set-up is the same as Fig.~\ref{fig:synthetic} except $Y_n \sim 10^{18}$ and $T_i \sim 15$~keV.}
    \label{fig:synthetic2}
\end{figure}

Modulations have also been observed in experiments with $Y_n\sim10^{18}$, $T_i \sim 15$~keV and foil configuration $(56.7 \ \mathrm{\mu m},\ 0.5\ \mathrm{cm^2})$. Repeating the previous analysis for this scenario leads to Fig.~\ref{fig:synthetic2}, where the errors introduced by energy modulations remain largely acceptable, except for one outlier circled in red. As $Y_n$ increases further, the impact of modulations will eventually become non-negligible. Here, the potential approaches to mitigate this effect are discussed.

One proposed solution is to move the foil back by an additional $\sim 50$~cm, which would reduce the neutron flux by a factor of $\sim 10$ and similarly lower $n_d$. While this would decrease the amplitude of modulations, it would not eliminate them entirely. Photo-ionized electrons can still populate the shadowed region before neutrons reach the foil, allowing modulations to persist. A more permanent solution as proposed by \cite{wink2024next} is to position the foil close to the MRS aperture. In this case, photo-ionized electrons would not have an opportunity to interact with deuterons, preventing the onset of TSI and consequently energy modulations.

\section{Conclusion}

To summarize, this work indicates that energy modulations observed in the MRS data obtained at the NIF are driven by the two-stream instability between deuterons and ambient electrons. The modulation features obtained in shot N240504-001 are reasonably reproduced by PIC simulations. A realistic density scan reveals that modulation amplitude scales linearly with the deuteron-to-electron density ratio. Synthetic data analysis based on this scaling suggests that the errors induced by modulations remain within the acceptable limit at $Y_n \sim 10^{17} - 10^{18}$. However, these effects are expected to worsen at higher yields. One viable solution is to place the foil directly at the MRS, which would eliminate the interference from photo-ionized electrons.

\section*{Acknowledgements}
We are grateful to Scott Vonhof (LLNL) for useful information about the diagnostic windows and Stuart Morris (University of Warwick) for his assistance with the \textsc{epoch} input deck. We also want to thank Wendi Sweet, Michael Weir (General Atomics) and Edwin Casco (LLNL) for their help in measuring the foil thickness post-exposure. 

\section*{Funding}

This work was performed under the auspices of the U.S. Department of Energy by LLNL under Contract DE-AC52-07NA27344. Additional financial support was provided by LLNL under Contract B656484, the NNSA MIT Center of Excellence under Contract DE-NA0003868, and the Laboratory of Laser Energetics under NNSA Contract DE-NA0004144. \textsc{epoch} was in part funded by the UK EPSRC grants EP/G054950/1, EP/G056803/1, EP/G055165/1 and EP/ M022463/1. The PIC simulations in this work were performed on the MIT-PSFC Engaging cluster funded by DOE grant number DE-FG02-91-ER54109. 

This report was prepared as an account of work sponsored by an agency of the United States Government. Neither the United States Government nor any agency thereof, nor any of their employees, makes any warranty, express or implied, or assumes any legal liability or responsibility for the accuracy, completeness, or usefulness of any information, apparatus, product, or process disclosed, or represents that its use would not infringe privately owned rights. Reference herein to any specific commercial product, process, or service by trade name, trademark, manufacturer, or otherwise does not necessarily constitute or imply its endorsement, recommendation, or favoring by the United States Government or any agency thereof. The views and opinions of authors expressed herein do not necessarily state or reflect those of the United States Government or any agency thereof.

\section*{Declaration of interests}

The authors report no conflict of interest.

\bibliographystyle{jpp}
\bibliography{jpp-instructions}

\begin{thebibliography}{50}
\expandafter\ifx\csname natexlab\endcsname\relax\def\natexlab#1{#1}\fi
\def\au#1{#1} \def\ed#1{#1} \def\yr#1{#1}\def\at#1{#1}\def\jt#1{\textit{#1}} \def\bt#1{#1}\def\bvol#1{\textbf{#1}} \def\vol#1{#1} \def\pg#1{#1} \def\publ#1{#1}\def\arxiv#1{#1}\def\org#1{#1}\def\st#1{\textit{#1}}

\bibitem[Abu-Shawareb {\em et~al.\/}(2022)Abu-Shawareb, Acree, Adams, Adams, Addis, Aden, Adrian, Afeyan, Aggleton, Aghaian {\em et~al.\/}]{abu2022lawson}
{\sc \au{Abu-Shawareb, H}, \au{Acree, R}, \au{Adams, P}, \au{Adams, J}, \au{Addis, B}, \au{Aden, R}, \au{Adrian, 2P}, \au{Afeyan, BB}, \au{Aggleton, M}, \au{Aghaian, L} \& \au{others}} \yr{2022}  \at{Lawson criterion for ignition exceeded in an inertial fusion experiment}.  \jt{Physical review letters}  \bvol{129}~(7),  \pg{075001}.

\bibitem[Agostinelli {\em et~al.\/}(2003)Agostinelli, Allison, Amako, Apostolakis, Araujo, Arce, Asai, Axen, Banerjee, Barrand {\em et~al.\/}]{agostinelli2003geant4}
{\sc \au{Agostinelli, Sea}, \au{Allison, John}, \au{Amako, K~al}, \au{Apostolakis, John}, \au{Araujo, Henrique}, \au{Arce, Pedro}, \au{Asai, Makoto}, \au{Axen, D}, \au{Banerjee, Swagato}, \au{Barrand, GJNI} \& \au{others}} \yr{2003}  \at{Geant4—a simulation toolkit}.  \jt{Nuclear instruments and methods in physics research section A: Accelerators, Spectrometers, Detectors and Associated Equipment}  \bvol{506}~(3),  \pg{250--303}.

\bibitem[Allen {\em et~al.\/}(2003)Allen, Sentoku, Audebert, Blazevic, Cowan, Fuchs, Gauthier, Geissel, Hegelich, Karsch {\em et~al.\/}]{allen2003proton}
{\sc \au{Allen, M}, \au{Sentoku, Y}, \au{Audebert, P}, \au{Blazevic, A}, \au{Cowan, T}, \au{Fuchs, J}, \au{Gauthier, JC}, \au{Geissel, M}, \au{Hegelich, Manuel}, \au{Karsch, Stefan} \& \au{others}} \yr{2003}  \at{Proton spectra from ultraintense laser--plasma interaction with thin foils: Experiments, theory, and simulation}.  \jt{Physics of Plasmas}  \bvol{10}~(8),  \pg{3283--3289}.

\bibitem[Allison {\em et~al.\/}(2006)Allison, Amako, Apostolakis, Araujo, Dubois, Asai, Barrand, Capra, Chauvie, Chytracek {\em et~al.\/}]{allison2006geant4}
{\sc \au{Allison, John}, \au{Amako, Katsuya}, \au{Apostolakis, JEA}, \au{Araujo, HAAH}, \au{Dubois, P~Arce}, \au{Asai, MAAM}, \au{Barrand, GABG}, \au{Capra, RACR}, \au{Chauvie, SACS}, \au{Chytracek, RACR} \& \au{others}} \yr{2006}  \at{{Geant4} developments and applications}.  \jt{IEEE Transactions on nuclear science}  \bvol{53}~(1),  \pg{270--278}.

\bibitem[Allison {\em et~al.\/}(2016)Allison, Amako, Apostolakis, Arce, Asai, Aso, Bagli, Bagulya, Banerjee, Barrand {\em et~al.\/}]{allison2016recent}
{\sc \au{Allison, John}, \au{Amako, Katsuya}, \au{Apostolakis, John}, \au{Arce, Pedro}, \au{Asai, Makoto}, \au{Aso, Tsukasa}, \au{Bagli, Enrico}, \au{Bagulya, A}, \au{Banerjee, S}, \au{Barrand, GJNI} \& \au{others}} \yr{2016}  \at{Recent developments in {Geant4}}.  \jt{Nuclear instruments and methods in physics research section A: Accelerators, Spectrometers, Detectors and Associated Equipment}  \bvol{835},  \pg{186--225}.

\bibitem[Arber {\em et~al.\/}(2015)Arber, Bennett, Brady, Lawrence-Douglas, Ramsay, Sircombe, Gillies, Evans, Schmitz, Bell {\em et~al.\/}]{arber2015contemporary}
{\sc \au{Arber, TD}, \au{Bennett, Keith}, \au{Brady, CS}, \au{Lawrence-Douglas, A}, \au{Ramsay, MG}, \au{Sircombe, Nathan~John}, \au{Gillies, Paddy}, \au{Evans, RG}, \au{Schmitz, Holger}, \au{Bell, AR} \& \au{others}} \yr{2015}  \at{Contemporary particle-in-cell approach to laser-plasma modelling}.  \jt{Plasma Physics and Controlled Fusion}  \bvol{57}~(11),  \pg{113001}.

\bibitem[Atzeni \& Meyer-ter Vehn(2004)]{atzeni2004physics}
{\sc \au{Atzeni, Stefano} \& \au{Meyer-ter Vehn, J{\"u}rgen}} \yr{2004} {\em The physics of inertial fusion: beam plasma interaction, hydrodynamics, hot dense matter\/}, ,  \vol{vol. 125}.  \publ{OUP Oxford}.

\bibitem[Ballabio {\em et~al.\/}(1998)Ballabio, K{\"a}llne \& Gorini]{ballabio1998relativistic}
{\sc \au{Ballabio, Luigi}, \au{K{\"a}llne, Jan} \& \au{Gorini, Giuseppe}} \yr{1998}  \at{Relativistic calculation of fusion product spectra for thermonuclear plasmas}.  \jt{Nuclear fusion}  \bvol{38}~(11),  \pg{1723}.

\bibitem[Bennett {\em et~al.\/}(2017)Bennett, Brady, Schmitz, Ridgers, Arber, Evans \& Bell]{bennett2017users}
{\sc \au{Bennett, Keith}, \au{Brady, Chris}, \au{Schmitz, Holger}, \au{Ridgers, Christopher}, \au{Arber, Tony}, \au{Evans, Roger} \& \au{Bell, Tony}} \yr{2017}  \at{Users manual for the {EPOCH PIC} codes}.  \jt{University of Warwick} .

\bibitem[Brown {\em et~al.\/}(2018)Brown, Chadwick, Capote, Kahler, Trkov, Herman, Sonzogni, Danon, Carlson, Dunn {\em et~al.\/}]{brown2018endf}
{\sc \au{Brown, David~A}, \au{Chadwick, Mark~Benjamin}, \au{Capote, R}, \au{Kahler, AC}, \au{Trkov, A}, \au{Herman, MW}, \au{Sonzogni, AA}, \au{Danon, Y}, \au{Carlson, AD}, \au{Dunn, M} \& \au{others}} \yr{2018}  \at{Endf/b-viii. 0: the 8th major release of the nuclear reaction data library with cielo-project cross sections, new standards and thermal scattering data}.  \jt{Nuclear Data Sheets}  \bvol{148},  \pg{1--142}.

\bibitem[Busch {\em et~al.\/}(2004)Busch, Shiryaev, Ter-Avetisyan, Schn{\"u}rer, Nickles \& Sandner]{busch2004shape}
{\sc \au{Busch, S}, \au{Shiryaev, O}, \au{Ter-Avetisyan, S}, \au{Schn{\"u}rer, M}, \au{Nickles, PV} \& \au{Sandner, W}} \yr{2004}  \at{Shape of ion energy spectra in ultra-short and intense laser--matter interaction}.  \jt{Applied Physics B}  \bvol{78},  \pg{911--914}.

\bibitem[Casey {\em et~al.\/}(2013)Casey, Frenje, Gatu~Johnson, S{\'e}guin, Li, Petrasso, Glebov, Katz, Magoon, Meyerhofer {\em et~al.\/}]{casey2013magnetic}
{\sc \au{Casey, D~T}, \au{Frenje, J.A}, \au{Gatu~Johnson, M}, \au{S{\'e}guin, FH}, \au{Li, CK}, \au{Petrasso, RD}, \au{Glebov, V~Yu}, \au{Katz, J}, \au{Magoon, J}, \au{Meyerhofer, DD} \& \au{others}} \yr{2013}  \at{The magnetic recoil spectrometer for measurements of the absolute neutron spectrum at {OMEGA} and the {NIF}}.  \jt{Review of Scientific Instruments}  \bvol{84}~(4),  \pg{043506}.

\bibitem[Casey {\em et~al.\/}(2011)Casey, Frenje, S{\'e}guin, Li, Rosenberg, Rinderknecht, Manuel, Gatu~Johnson, Schaeffer, Frankel {\em et~al.\/}]{casey2011coincidence}
{\sc \au{Casey, D~T}, \au{Frenje, J.A}, \au{S{\'e}guin, FH}, \au{Li, CK}, \au{Rosenberg, MJ}, \au{Rinderknecht, H}, \au{Manuel, MJ-E}, \au{Gatu~Johnson, M}, \au{Schaeffer, JC}, \au{Frankel, R} \& \au{others}} \yr{2011}  \at{The coincidence counting technique for orders of magnitude background reduction in data obtained with the magnetic recoil spectrometer at omega and the {NIF}}.  \jt{Review of scientific instruments}  \bvol{82}~(7),  \pg{073502}.

\bibitem[Chen(2012)]{chen2012introduction}
{\sc \au{Chen, Francis~F}} \yr{2012} {\em Introduction to plasma physics\/}.  \publ{Springer Science \& Business Media}, chap. 6.

\bibitem[Chen {\em et~al.\/}(2025)Chen, Woods, Farmer, Aybar, Liedahl, MacLaren, Schneider, Scott, Harte, Hinkel, Landen, Moody, Rosen, Ross, Rogers, Roskopf, Swadling, Vonhof \& Zimmerman]{chen2025drive}
{\sc \au{Chen, Hui}, \au{Woods, D.~T.}, \au{Farmer, W.~A.}, \au{Aybar, N.~A.}, \au{Liedahl, D.~A.}, \au{MacLaren, S.~A.}, \au{Schneider, M.~B.}, \au{Scott, H.~A.}, \au{Harte, J.~A.}, \au{Hinkel, D.~E.}, \au{Landen, O.~L.}, \au{Moody, J.~D.}, \au{Rosen, M.~D.}, \au{Ross, J.~S.}, \au{Rogers, S.}, \au{Roskopf, N.}, \au{Swadling, G.}, \au{Vonhof, S.} \& \au{Zimmerman, G.~B.}} \yr{2025}  \at{Key advancements toward eliminating the “drive deficit” in {ICF} hohlraum simulations}.  \jt{Physics of Plasmas}  \bvol{32}~(4),  \pg{042704}.

\bibitem[Davidson {\em et~al.\/}(1972)Davidson, Hammer, Haber \& Wagner]{davidson1972nonlinear}
{\sc \au{Davidson, Ronald~C}, \au{Hammer, David~A}, \au{Haber, Irving} \& \au{Wagner, Carl~E}} \yr{1972}  \at{Nonlinear development of electromagnetic instabilities in anisotropic plasmas}.  \jt{The Physics of Fluids}  \bvol{15}~(2),  \pg{317--333}.

\bibitem[Decker {\em et~al.\/}(1997)Decker, Turner, Landen, Suter, Amendt, Kornblum, Hammel, Murphy, Wallace, Delamater {\em et~al.\/}]{decker1997hohlraum}
{\sc \au{Decker, C}, \au{Turner, RE}, \au{Landen, OL}, \au{Suter, LJ}, \au{Amendt, P}, \au{Kornblum, HN}, \au{Hammel, BA}, \au{Murphy, TJ}, \au{Wallace, J}, \au{Delamater, ND} \& \au{others}} \yr{1997}  \at{Hohlraum radiation drive measurements on the {Omega} laser}.  \jt{Physical review letters}  \bvol{79}~(8),  \pg{1491}.

\bibitem[Dewald {\em et~al.\/}(2004)Dewald, Campbell, Turner, Holder, Landen, Glenzer, Kauffman, Suter, Landon, Rhodes {\em et~al.\/}]{dewald2004dante}
{\sc \au{Dewald, EL}, \au{Campbell, KM}, \au{Turner, RE}, \au{Holder, JP}, \au{Landen, OL}, \au{Glenzer, SH}, \au{Kauffman, RL}, \au{Suter, LJ}, \au{Landon, M}, \au{Rhodes, M} \& \au{others}} \yr{2004}  \at{{Dante soft x-ray power diagnostic for National Ignition Facility}}.  \jt{Review of Scientific Instruments}  \bvol{75}~(10),  \pg{3759--3761}.

\bibitem[Dewald {\em et~al.\/}(2020)Dewald, Landen, Salmonson, Masse, Tabak, Smalyuk, Schiaffino, Heredia, Schneider \& Nikroo]{dewald2020first}
{\sc \au{Dewald, EL}, \au{Landen, OL}, \au{Salmonson, J}, \au{Masse, L}, \au{Tabak, M}, \au{Smalyuk, VA}, \au{Schiaffino, S}, \au{Heredia, R}, \au{Schneider, M} \& \au{Nikroo, A}} \yr{2020}  \at{{First study of Hohlraum x-ray preheat asymmetry inside an ICF capsule}}.  \jt{Physics of Plasmas}  \bvol{27}~(12),  \pg{122703}.

\bibitem[Frenje {\em et~al.\/}(2010)Frenje, Casey, Li, S{\'e}guin, Petrasso, Glebov, Radha, Sangster, Meyerhofer, Hatchett {\em et~al.\/}]{frenje2010probing}
{\sc \au{Frenje, J.A}, \au{Casey, D~T}, \au{Li, CK}, \au{S{\'e}guin, FH}, \au{Petrasso, RD}, \au{Glebov, V~Yu}, \au{Radha, PB}, \au{Sangster, TC}, \au{Meyerhofer, DD}, \au{Hatchett, SP} \& \au{others}} \yr{2010}  \at{Probing high areal-density cryogenic deuterium-tritium implosions using downscattered neutron spectra measured by the magnetic recoil spectrometer}.  \jt{Physics of Plasmas}  \bvol{17}~(5),  \pg{056311}.

\bibitem[Frenje(2020)]{frenje2020nuclear}
{\sc \au{Frenje, J.~A}} \yr{2020}  \at{Nuclear diagnostics for inertial confinement fusion {(ICF)} plasmas}.  \jt{Plasma Physics and Controlled Fusion}  \bvol{62}~(2),  \pg{023001}.

\bibitem[Frenje {\em et~al.\/}(2013)Frenje, Bionta, Bond, Caggiano, Casey, Cerjan, Edwards, Eckart, Fittinghoff, Friedrich {\em et~al.\/}]{frenje2013diagnosing}
{\sc \au{Frenje, Johan~A}, \au{Bionta, R}, \au{Bond, EJ}, \au{Caggiano, JA}, \au{Casey, DT}, \au{Cerjan, C}, \au{Edwards, J}, \au{Eckart, M}, \au{Fittinghoff, DN}, \au{Friedrich, S} \& \au{others}} \yr{2013}  \at{Diagnosing implosion performance at the {National Ignition Facility (NIF)} by means of neutron spectrometry}.  \jt{Nuclear Fusion}  \bvol{53}~(4),  \pg{043014}.

\bibitem[Gatu~Johnson {\em et~al.\/}(2022)Gatu~Johnson, Johnson, Lahmann, S{\'e}guin, Sperry, Bhandarkar, Bionta, Casco, Casey, Mackinnon {\em et~al.\/}]{gatu2022high}
{\sc \au{Gatu~Johnson, M}, \au{Johnson, TM}, \au{Lahmann, BJ}, \au{S{\'e}guin, FH}, \au{Sperry, B}, \au{Bhandarkar, N}, \au{Bionta, RM}, \au{Casco, E}, \au{Casey, DT}, \au{Mackinnon, AJ} \& \au{others}} \yr{2022}  \at{{High-yield magnetic recoil neutron spectrometer on the National Ignition Facility for operation up to 60 MJ}}.  \jt{Review of Scientific Instruments}  \bvol{93}~(8),  \pg{083513}.

\bibitem[Hara {\em et~al.\/}(2018)Hara, Kaganovich \& Startsev]{hara2018generation}
{\sc \au{Hara, Kentaro}, \au{Kaganovich, Igor~D} \& \au{Startsev, Edward~A}} \yr{2018}  \at{Generation of forerunner electron beam during interaction of ion beam pulse with plasma}.  \jt{Physics of Plasmas}  \bvol{25}~(1),  \pg{011609}.

\bibitem[Hicks {\em et~al.\/}(2001)Hicks, Li, S{\'e}guin, Schnittman, Ram, Frenje, Petrasso, Soures, Meyerhofer, Roberts {\em et~al.\/}]{hicks2001observations}
{\sc \au{Hicks, D.~G}, \au{Li, CK}, \au{S{\'e}guin, FH}, \au{Schnittman, JD}, \au{Ram, AK}, \au{Frenje, J.~A}, \au{Petrasso, RD}, \au{Soures, JM}, \au{Meyerhofer, DD}, \au{Roberts, S} \& \au{others}} \yr{2001}  \at{Observations of fast protons above 1 mev produced in direct-drive laser-fusion experiments}.  \jt{Physics of Plasmas}  \bvol{8}~(2),  \pg{606--610}.

\bibitem[Hohenberger {\em et~al.\/}(2013)Hohenberger, Palmer, LaCaille, Dewald, Divol, Bond, D{\"o}ppner, Lee, Kauffman, Salmonson {\em et~al.\/}]{hohenberger2013measuring}
{\sc \au{Hohenberger, M}, \au{Palmer, NE}, \au{LaCaille, G}, \au{Dewald, EL}, \au{Divol, L}, \au{Bond, EJ}, \au{D{\"o}ppner, T}, \au{Lee, JJ}, \au{Kauffman, RL}, \au{Salmonson, JD} \& \au{others}} \yr{2013} Measuring the hot-electron population using time-resolved hard x-ray detectors on the {NIF}.  \bt{In {\em Target Diagnostics Physics and Engineering for Inertial Confinement Fusion II\/}}, ,  \vol{vol. 8850},  \pg{pp. 106--114}. SPIE.

\bibitem[Hou {\em et~al.\/}(2015)Hou, Chen, Yu \& Wu]{hou2015linear}
{\sc \au{Hou, YW}, \au{Chen, MX}, \au{Yu, MY} \& \au{Wu, B}} \yr{2015}  \at{Linear and nonlinear behaviour of two-stream instabilities in collisionless plasmas}.  \jt{Journal of Plasma Physics}  \bvol{81}~(6),  \pg{905810602}.

\bibitem[Hu {\em et~al.\/}(2013)Hu, Song, Zhao \& Wang]{hu2013modulation}
{\sc \au{Hu, Zhang-Hu}, \au{Song, Yuan-Hong}, \au{Zhao, Yong-Tao} \& \au{Wang, You-Nian}} \yr{2013}  \at{Modulation of continuous ion beams with low drift velocity by induced wakefield in background plasmas}.  \jt{Laser and Particle Beams}  \bvol{31}~(1),  \pg{135--140}.

\bibitem[Hui {\em et~al.\/}(2019)Hui, Hu, Zhao, Cheng, Mei \& Wang]{hui2019modulation}
{\sc \au{Hui, De-Xuan}, \au{Hu, Zhang-Hu}, \au{Zhao, Yong-Tao}, \au{Cheng, Rui}, \au{Mei, Xian-Xiu} \& \au{Wang, You-Nian}} \yr{2019}  \at{Modulation of ion beams in two-component plasmas: Three-dimensional particle-in-cell simulation}.  \jt{Physics of Plasmas}  \bvol{26}~(9),  \pg{093104}.

\bibitem[Khan {\em et~al.\/}(2018)Khan, Jarrott, Patel, Izumi, Ma, MacPhee, Hatch, Landen, Heinmiller, Kilkenny {\em et~al.\/}]{khan2018implementing}
{\sc \au{Khan, SF}, \au{Jarrott, LC}, \au{Patel, PK}, \au{Izumi, N}, \au{Ma, T}, \au{MacPhee, AG}, \au{Hatch, B}, \au{Landen, OL}, \au{Heinmiller, J}, \au{Kilkenny, JD} \& \au{others}} \yr{2018}  \at{Implementing time resolved electron temperature capability at the {NIF} using a streak camera}.  \jt{Review of Scientific Instruments}  \bvol{89}~(10),  \pg{10K117}.

\bibitem[Lemmon {\em et~al.\/}(2010)Lemmon, Huber, McLinden {\em et~al.\/}]{lemmon2010nist}
{\sc \au{Lemmon, Eric~W}, \au{Huber, Marcia~L}, \au{McLinden, Mark~O} \& \au{others}} \yr{2010}  \at{Nist standard reference database 23}.  \jt{Reference fluid thermodynamic and transport properties (REFPROP), version}  \bvol{9}.

\bibitem[Lindl {\em et~al.\/}(2014)Lindl, Landen, Edwards \& Moses]{lindl2014review}
{\sc \au{Lindl, John}, \au{Landen, Otto}, \au{Edwards, John} \& \au{Moses}} \yr{2014}  \at{Review of the national ignition campaign 2009-2012}.  \jt{Physics of Plasmas}  \bvol{21}~(2),  \pg{020501}.

\bibitem[Lindl {\em et~al.\/}(2004)Lindl, Amendt, Berger, Glendinning, Glenzer, Haan, Kauffman, Landen \& Suter]{lindl2004physics}
{\sc \au{Lindl, John~D}, \au{Amendt, Peter}, \au{Berger, Richard~L}, \au{Glendinning, S~Gail}, \au{Glenzer, Siegfried~H}, \au{Haan, Steven~W}, \au{Kauffman, Robert~L}, \au{Landen, Otto~L} \& \au{Suter, Laurence~J}} \yr{2004}  \at{The physics basis for ignition using indirect-drive targets on the {National Ignition Facility}}.  \jt{Physics of plasmas}  \bvol{11}~(2),  \pg{339--491}.

\bibitem[Lv {\em et~al.\/}(2023)Lv, Wang, Liu, Li, Cheng, Huang, Li, Zhang, Chen, Wang {\em et~al.\/}]{lv2023ion}
{\sc \au{Lv, SY}, \au{Wang, Qing}, \au{Liu, DJ}, \au{Li, XX}, \au{Cheng, RJ}, \au{Huang, ZM}, \au{Li, XM}, \au{Zhang, ST}, \au{Chen, ZJ}, \au{Wang, Qiang} \& \au{others}} \yr{2023}  \at{The ion acoustic instability during collisionless two-ion-species plasma expansion}.  \jt{Physics of Plasmas}  \bvol{30}~(9),  \pg{092113}.

\bibitem[Marinak {\em et~al.\/}(2001)Marinak, Kerbel, Gentile, Jones, Munro, Pollaine, Dittrich \& Haan]{marinak2001three}
{\sc \au{Marinak, Michael~M}, \au{Kerbel, GD}, \au{Gentile, NA}, \au{Jones, O}, \au{Munro, D}, \au{Pollaine, S}, \au{Dittrich, TR} \& \au{Haan, SW}} \yr{2001}  \at{Three-dimensional {HYDRA simulations of National Ignition Facility targets}}.  \jt{Physics of Plasmas}  \bvol{8}~(5),  \pg{2275--2280}.

\bibitem[May {\em et~al.\/}(2015)May, Fournier, Colvin, Barrios, Dewald, Hohenberger, Moody, Patterson, Schneider, Widmann {\em et~al.\/}]{may2015bright}
{\sc \au{May, MJ}, \au{Fournier, KB}, \au{Colvin, JD}, \au{Barrios, MA}, \au{Dewald, EL}, \au{Hohenberger, M}, \au{Moody, J}, \au{Patterson, JR}, \au{Schneider, M}, \au{Widmann, K} \& \au{others}} \yr{2015}  \at{Bright x-ray stainless steel k-shell source development at the {National Ignition Facility}}.  \jt{Physics of Plasmas}  \bvol{22}~(6),  \pg{063305}.

\bibitem[McMillan(2020)]{mcmillan2020necessary}
{\sc \au{McMillan, BF}} \yr{2020}  \at{Is it necessary to resolve the debye length in standard or $\delta$f {PIC} codes?}  \jt{Physics of Plasmas}  \bvol{27}~(5),  \pg{052106}.

\bibitem[Milovich {\em et~al.\/}(2020)Milovich, Casey, MacGowan, Clark, Mariscal, Ma, Baker, Bionta, Hahn, Moore {\em et~al.\/}]{milovich2020understanding}
{\sc \au{Milovich, JL}, \au{Casey, DC}, \au{MacGowan, B}, \au{Clark, D}, \au{Mariscal, D}, \au{Ma, T}, \au{Baker, K}, \au{Bionta, R}, \au{Hahn, K}, \au{Moore, A} \& \au{others}} \yr{2020}  \at{Understanding asymmetries using integrated simulations of capsule implosions in low gas-fill hohlraums at the {National Ignition Facility}}.  \jt{Plasma Physics and Controlled Fusion}  \bvol{63}~(2),  \pg{025012}.

\bibitem[Milovich {\em et~al.\/}(2010)Milovich, Dewald, Thomas, Robey \& Landen]{milovich2010tuning}
{\sc \au{Milovich, J}, \au{Dewald, E}, \au{Thomas, C}, \au{Robey, H} \& \au{Landen, O}} \yr{2010} Tuning experiments for the first 2 ns of the {NIF} ignition pulse.  \bt{In {\em APS Division of Plasma Physics Meeting Abstracts\/}}, ,  \vol{vol.~52},  \pg{pp. NO5--007}.

\bibitem[Mima {\em et~al.\/}(2018)Mima, Fuchs, Taguchi, Alvarez, Marqu{\`e}s, Chen, Tajima \& Perlado]{mima2018self}
{\sc \au{Mima, K}, \au{Fuchs, J}, \au{Taguchi, T}, \au{Alvarez, J}, \au{Marqu{\`e}s, JR}, \au{Chen, SN}, \au{Tajima, T} \& \au{Perlado, JM}} \yr{2018}  \at{Self-modulation and anomalous collective scattering of laser produced intense ion beam in plasmas}.  \jt{Matter and Radiation at Extremes}  \bvol{3}~(3),  \pg{127--134}.

\bibitem[Reid(2003)]{reid2003photoelectron}
{\sc \au{Reid, Katharine~L.}} \yr{2003}  \at{Photoelectron angular distributions}.  \jt{Annual Review of Physical Chemistry}  \bvol{54},  \pg{397--424}.

\bibitem[Rinderknecht(2015)]{rinderknecht2015studies}
{\sc \au{Rinderknecht, Hans~G}} \yr{2015}  \at{Studies of non-hydrodynamic processes in icf implosions on {OMEGA} and the {National Ignition Facility}}. PhD thesis, Massachusetts Institute of Technology.

\bibitem[Smalyuk {\em et~al.\/}(2014)Smalyuk, Tipton, Pino, Casey, Grim, Remington, Rowley, Weber, Barrios, Benedetti {\em et~al.\/}]{smalyuk2014measurements}
{\sc \au{Smalyuk, VA}, \au{Tipton, RE}, \au{Pino, JE}, \au{Casey, DT}, \au{Grim, GP}, \au{Remington, BA}, \au{Rowley, DP}, \au{Weber, SV}, \au{Barrios, M}, \au{Benedetti, LR} \& \au{others}} \yr{2014}  \at{Measurements of an ablator-gas atomic mix in indirectly driven implosions at the {National Ignition Facility}}.  \jt{Physical review letters}  \bvol{112}~(2),  \pg{025002}.

\bibitem[Swadling {\em et~al.\/}(2017)Swadling, Ross, Manha, Galbraith, Datte, Sorce, Katz, Froula, Widmann, Jones {\em et~al.\/}]{swadling2017initial}
{\sc \au{Swadling, GF}, \au{Ross, JS}, \au{Manha, D}, \au{Galbraith, J}, \au{Datte, P}, \au{Sorce, C}, \au{Katz, J}, \au{Froula, DH}, \au{Widmann, K}, \au{Jones, OS} \& \au{others}} \yr{2017}  \at{Initial experimental demonstration of the principles of a xenon gas shield designed to protect optical components from soft x-ray induced opacity (blanking) in high energy density experiments}.  \jt{Physics of Plasmas}  \bvol{24}~(3),  \pg{032705}.

\bibitem[Takabe(2023)]{takabe2023theory}
{\sc \au{Takabe, Hideaki}} \yr{2023}  \at{Theory of magnetic turbulence and shock formation induced by a collisionless plasma instability}.  \jt{Physics of Plasmas}  \bvol{30}~(3),  \pg{030901}.

\bibitem[Tikhonchuk {\em et~al.\/}(2005)Tikhonchuk, Andreev, Bochkarev \& Bychenkov]{tikhonchuk2005ion}
{\sc \au{Tikhonchuk, VT}, \au{Andreev, AA}, \au{Bochkarev, SG} \& \au{Bychenkov, V~Yu}} \yr{2005}  \at{Ion acceleration in short-laser-pulse interaction with solid foils}.  \jt{Plasma physics and controlled fusion}  \bvol{47}~(12B),  \pg{B869}.

\bibitem[Virtanen {\em et~al.\/}(2020)Virtanen, Gommers, Oliphant, Haberland, Reddy, Cournapeau, Burovski, Peterson, Weckesser, Bright {\em et~al.\/}]{virtanen2020scipy}
{\sc \au{Virtanen, Pauli}, \au{Gommers, Ralf}, \au{Oliphant, Travis~E}, \au{Haberland, Matt}, \au{Reddy, Tyler}, \au{Cournapeau, David}, \au{Burovski, Evgeni}, \au{Peterson, Pearu}, \au{Weckesser, Warren}, \au{Bright, Jonathan} \& \au{others}} \yr{2020}  \at{Scipy 1.0: fundamental algorithms for scientific computing in python}.  \jt{Nature methods}  \bvol{17}~(3),  \pg{261--272}.

\bibitem[Wink {\em et~al.\/}(2024)Wink, Gatu~Johnson, Mackie, Kunimune, Dannhoff, Lawrence, Berg, Casey, Schlossberg, Gopalaswamy {\em et~al.\/}]{wink2024next}
{\sc \au{Wink, C~W}, \au{Gatu~Johnson, M}, \au{Mackie, S}, \au{Kunimune, JH}, \au{Dannhoff, SG}, \au{Lawrence, Y}, \au{Berg, GPA}, \au{Casey, DT}, \au{Schlossberg, DJ}, \au{Gopalaswamy, V} \& \au{others}} \yr{2024}  \at{The next-generation magnetic recoil spectrometer ({MRSnext}) on {OMEGA} and {NIF} for diagnosing ion temperature, yield, areal density, and alpha heating}.  \jt{Review of Scientific Instruments}  \bvol{95}~(8),  \pg{083548}.

\bibitem[Yeh \& Lindau(1993)]{yeh1993atomic}
{\sc \au{Yeh, JJ} \& \au{Lindau, I}} \yr{1993}  \at{Atomic and nuclear data tables}.  \jt{AT \& T, Gordon Breach, Langhorne} .

\bibitem[Zylstra {\em et~al.\/}(2021)Zylstra, Kritcher, Hurricane, Callahan, Baker, Braun, Casey, Clark, Clark, D{\"o}ppner {\em et~al.\/}]{zylstra2021record}
{\sc \au{Zylstra, AB}, \au{Kritcher, AL}, \au{Hurricane, OA}, \au{Callahan, DA}, \au{Baker, K}, \au{Braun, T}, \au{Casey, DT}, \au{Clark, D}, \au{Clark, K}, \au{D{\"o}ppner, T} \& \au{others}} \yr{2021}  \at{Record energetics for an inertial fusion implosion at {NIF}}.  \jt{Physical review letters}  \bvol{126}~(2),  \pg{025001}.

\end{thebibliography}

\end{document}